\documentclass[%
 reprint,
 amsmath,amssymb,
 aps,
]{revtex4-2}

\usepackage[utf8]{inputenc}
\usepackage[T1]{fontenc}
\usepackage[english]{babel}
\usepackage{graphicx}
\usepackage{dcolumn}
\usepackage{bm}

\begin{document}

\preprint{APS/123-QED}

\title{Quantitative and Predictive Folding Models from Limited Single-Molecule Data Using Simulation-Based Inference}

\author{Lars Dingeldein$^{1, 2}$, Aaron Lyons$^{3}$, Pilar Cossio$^{4, 5}$, Michael Woodside$^{3, 6, 7}$, Roberto Covino$^{8,2}$}
\email{covino@fias.uni-frankfurt.de}

\affiliation{$^{1}$Institute of Physics, Goethe University Frankfurt, Frankfurt am Main, Germany}
\affiliation{$^{2}$Frankfurt Institute for Advanced Studies, Frankfurt am Main, Germany}
\affiliation{$^{3}$Department of Physics, University of Alberta, Edmont, Alberta, Canada}
\affiliation{$^{4}$Center for Computational Mathematics, Flatiron Institute, New York, United States}
\affiliation{$^{5}$Center for Computational Biology, Flatiron Institute, New York, United States}
\affiliation{$^{6}$Centre for Prions and Protein Folding Diseases, University of Alberta, Edmonton, Alberta, Canada}
\affiliation{$^{7}$Li Ka Shing Institute of Virology, University of Alberta, Edmonton, Alberta, Canada}
\affiliation{$^{8}$Institute of Computer science, Goethe University Frankfurt, Frankfurt am Main, Germany}

\date{\today}

\begin{abstract}
The study of biomolecular folding has been greatly advanced by single-molecule force spectroscopy (SMFS), which enables the observation of the dynamics of individual molecules. However, extracting quantitative models of fundamental properties such as folding landscapes from SMFS data is very challenging due to instrumental noise, linker artifacts, and the inherent stochasticity of the process, often requiring extensive datasets and complex calibration. Here, we introduce a framework based on simulation-based inference (SBI) that overcomes these limitations by integrating physics-based modeling with deep learning. We first apply this framework to analyze constant-force measurements of a DNA hairpin. From a single experimental trajectory of only two seconds, we successfully reconstruct the hairpin’s free energy landscape and folding dynamics, obtaining results in close agreement with established deconvolution methods that require 10–100 times more data. Furthermore, we demonstrate the generality of our approach by applying it to a riboswitch aptamer featuring multiple states and tertiary contacts, resolving the profile of a landscape featuring four metastable states from a single trajectory. The Bayesian nature of this approach robustly quantifies uncertainties for all inferred parameters, including diffusion coefficients and linker stiffness, without needing independent measurements of instrument properties. The inferred models are predictive, generating simulated trajectories that quantitatively reproduce experimental thermodynamics and kinetics. The ability to derive statistically robust models from minimal datasets is crucial for investigating complex biomolecular systems where extensive data collection is impractical, paving the way for novel applications of SMFS.
\end{abstract}

\maketitle

\noindent
Biomolecular folding is a fundamental process in which proteins, nucleic acids, and other biopolymers adopt three-dimensional structures essential for their function \cite{dill2012protein}. Understanding folding mechanisms is crucial to elucidating the principles of biomolecular self-organization and its connection to diseases. The folding dynamics arise from a complex interplay between many intermolecular forces, often modeled as a diffusive process on a rugged free energy landscape \cite{dill_levinthal_1997}. Characterizing the folding free energy landscape provides a complete thermodynamic picture of the folding process, providing insights into mechanisms and kinetics. However, experimentally obtaining free energy landscapes remains challenging due to limitations in spatial and temporal resolution \cite{buchner2005protein}.

Single-molecule force spectroscopy (SMFS) probes the folding dynamics of individual biomolecules held under tension  \cite{petrosyan_single-molecule_2021, neuman_single-molecule_2008}. These measurements enable the reconstruction of free energy landscapes along a one-dimensional reaction coordinate, usually the molecular extension \cite{hummer_free_2001,woodside_direct_2006,dudko_intrinsic_2006,dudko_theory_2008,hummer_free_2010}. Such reconstructions have been achieved using constant-force \cite{woodside_reconstructing_2014}, constant-position \cite{gebhardt2020full,walder_high-precision_2018}, and non-equilibrium measurements \cite{gupta_experimental_2011}.

\begin{figure*}[t]
    \centering
    \includegraphics[width=0.9\textwidth]{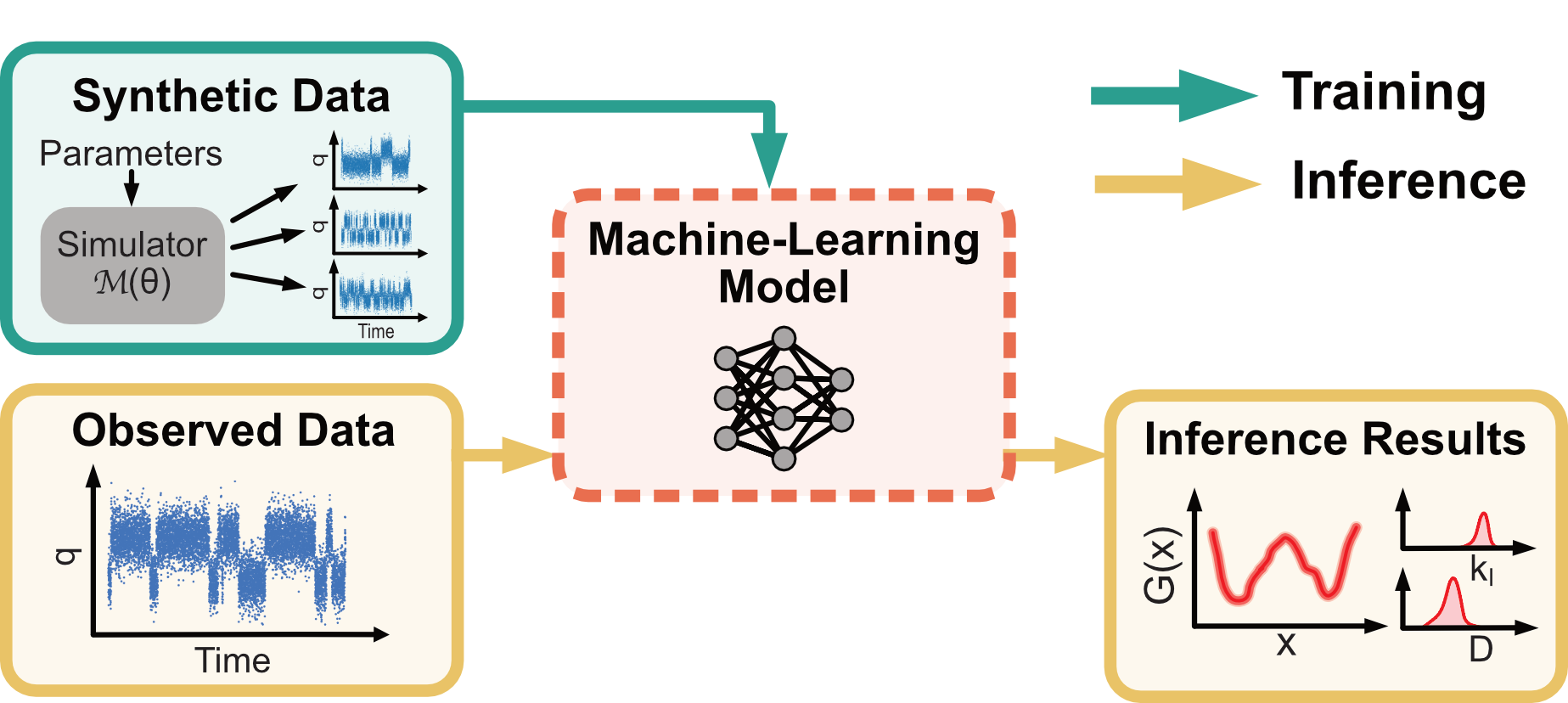}
    \caption{Framework for analyzing single-molecule force spectroscopy data using simulation-based inference. The process begins by generating simulated trajectories via a physics-based simulator (Top left). These trajectories are used to train a machine-learning model to establish probabilistic relationships between model parameters and synthetic data (Middle). The trained model is then evaluated using experimental data (Lower left), producing a distribution of model parameters that are compatible with the experimental observations (Lower right).}
    \label{fig:1}
\end{figure*}

However, SMFS measurements are indirect: the biomolecule is connected to a large pulling device through long, flexible linkers, so the measured extension reflects a convolution of molecular dynamics, linker fluctuations, and the instrument response \cite{cossio_artifacts_2015, neupane_quantifying_2016}. Measurement noise and the inherent stochasticity of single-molecule trajectories further complicate the estimation of the underlying free energy landscape. In principle, the linker contribution can be removed by deconvolution \cite{hinczewski_mechanical_2013, woodside_direct_2006, woodside_reconstructing_2014}. In practice, this requires a large amount of data and a very precise characterization of the linker, which is technically challenging and labor-intensive. Consequently, accurate free energy reconstructions exist only for a handful of molecules.

An alternative approach to overcome these problems is to reconstruct the free energy landscape by fitting a dynamical model $\mathcal{M}(\boldsymbol{\theta})$ to the observed trajectory $\boldsymbol{q}_{[1:N]}$, a vector containing $N$ measured extensions sampled at equal time intervals.  The parameters $\boldsymbol{\theta}$ determine the landscape, and must be inferred from the observed trajectory \cite{dingeldein_simulation-based_2022}. Bayesian inference provides a systematic statistical framework to determine the posterior distribution over $\boldsymbol{\theta}$ that is most consistent with $\boldsymbol{q}_{[1:N]}$:

\begin{align}
\label{eq:bayes}
    p(\boldsymbol{\theta}|\boldsymbol{q}_{[1:N]}) = \frac{p(\boldsymbol{q}_{[1:N]}|\boldsymbol{\theta})p(\boldsymbol{\theta})}{p(\boldsymbol{q}_{[1:N]})}
\end{align}
Here, $p(\boldsymbol{\theta})$ is the prior, encoding prior knowledge, $p(\boldsymbol{q}_{[1:N]}|\boldsymbol{\theta})$ is the likelihood of observing the trajectory $\boldsymbol{q}_{[1:N]}$ given $\boldsymbol{\theta}$, and $p(\boldsymbol{q}_{[1:N]})$ is a normalization factor.

Typical inference algorithms maximize the likelihood with respect to the parameters $\boldsymbol{\theta}$. However, evaluating the likelihood for partially observable dynamical models is very challenging, because it is a marginalized likelihood over all possible ``latent'' trajectories of the actual molecular extension $\boldsymbol{x}_{[1:N]}$ \cite{covino_2019, dingeldein_simulation-based_2025}:

\begin{align}
\label{eq: likelihood}
    p(\boldsymbol{q}_{[1:N]}|\boldsymbol{\theta}) = \int \mathcal{D}\boldsymbol{x}_{[1:N]} p(\boldsymbol{q}_{[1:N]}, \boldsymbol{x}_{[1:N]}|\boldsymbol{\theta})
\end{align}

An emerging approach to tackle this problem is simulation-based inference (SBI) \cite{cranmer_frontier_2020, dingeldein_simulation-based_2022,dingeldein_simulation-based_2025}, which integrates physics-based models with probabilistic deep learning to learn a surrogate model of the posterior \cite{papamakarios_fast_2018}, likelihood \cite{papamakarios_sequential_2019}, or likelihood ratio \cite{durkan_contrastive_2020}. The core idea is to use a simulator to generate synthetic measurements for specific parameters $\boldsymbol{\theta}$. Using deep neural networks, we can learn a surrogate model of the posterior distribution from the simulated data, providing a probabilistic mapping between parameters and data. SBI has shown success across various fields, from astrophysics \cite{dax_real-time_2021, regaldo-saint_blancard_galaxy_2024} to neuroscience \cite{gao_deep_2024, lueckmann_flexible_nodate}, particularly for models with complex likelihoods \cite{lueckmann_benchmarking_2021}.

In this work, we showcase how to leverage simulation-based inference to learn quantitative folding models from single-molecule force spectroscopy experiments by learning the Bayesian posterior from synthetic data. This framework yields accurate uncertainty estimates and quantitative predictive models, providing an efficient and robust approach for analyzing SMFS data. We applied it to the well-characterized 30R50/T4 DNA hairpin as a model system \cite{neupane2015transition, neupane2012transition, manuel2015reconstructing, neupane_quantifying_2016}, and showed that it can reconstruct the hairpin’s free energy landscape even from limited experimental data. We further demonstrated the generality of the approach by applying it to the multi-state folding of a riboswitch aptamer, highlighting its applicability to more complex systems.\\\\

\noindent
Single nucleic acid hairpins fold into simple structures, offering a well-defined and biologically relevant experimental framework for investigating folding mechanisms \cite{liphardt_reversible_2001, varani_exceptionally_1995, davydova_bacteriophage_2007}.
The end-to-end distance of the DNA serves as a convenient reaction coordinate for the folding reaction, along which the dynamics are well characterized by one-dimensional Brownian diffusion on an effective free energy landscape \cite{neupane2015transition, neupane_protein_2016}. 

We applied our inference framework \cite{dingeldein_simulation-based_2022} to learn a quantitative folding model from constant-force measurements of a 30R50/T4 DNA hairpin. This particular hairpin has been extensively studied, with measurements of the folding dynamics spanning from microseconds to minutes and folding energy landscape reconstructions using various methods and experimental setups \cite{woodside_direct_2006,engel2014prl,neupane2015transition,gupta_experimental_2011, lyons_quantifying_2024}.

The physical model $\mathcal{M}(\boldsymbol{\theta})$ describes both the biomolecular folding process and the measurement of the observed data. Given a measured trajectory $\boldsymbol{q}_{[1:N]}$, we aim to determine the optimal parameter set $\boldsymbol{\theta}$ that best fits the measurement by computing the posterior $p(\boldsymbol{\theta}|\boldsymbol{q}_{[1:N]}$). 

The harmonic-spring model $\mathcal{M}(\boldsymbol{\theta})$ provides a well-established framework for describing single-molecule force spectroscopy experiments \cite{covino_2019, cossio_artifacts_2015, hummer_free_2010}. It describes the coupled dynamics of the biomolecule and the measuring apparatus as a diffusive process on a two-dimensional free energy surface $G(q, x)$. Here, $q$ is the measured extension, including both the linker and the molecule, while $x$ is the (hidden) molecular extension. The free energy surface decomposes as $G(q, x) = G_0(x) + 1/2\cdot k_l(q-x)^2$ where $G_0(x)$ is the intrinsic folding free energy landscape of interest. The second term describes the linker coupling to the apparatus. We assume anisotropic Brownian dynamics on this energy landscape, with diffusion coefficient $D_x$ for the molecule and $D_q$ for the apparatus. The molecular free energy profile $G_0(x)$ is modeled using cubic spline interpolation. The model parameters $\boldsymbol{\theta} = \{D_q/D_x, k_l, G_0(x_0), G_0(x_1), ..., G_0(x_{N_\textrm{node}})\}$ are the ratio of diffusion coefficients $D_q/D_x$, the linker stiffness $k_l$, and $N_{\textrm{node}}$ positions of the spline nodes $G_0(x_i)$ to characterize the free energy profile $G_0(x)$. Inferring $\boldsymbol{\theta}$ from a measured trajectory $\boldsymbol{q}_{[1:N]}$ fully characterizes biomolecular folding as diffusion on the one-dimensional landscape $G_0(x)$.

Direct computation of the likelihood (Eq. \ref{eq: likelihood}) is extremely challenging because it requires integrating over all hidden molecular trajectories $\boldsymbol{x}_{[1:N]}$. In other words, computing the likelihood involves searching over all possible paths $\boldsymbol{x}_{[1:N]}$ consistent with the observed trajectory $\boldsymbol{q}_{[1:N]}$. In contrast, simulating a measured trajectory $\boldsymbol{q}_{[1:N]} \sim p(\boldsymbol{q}_{[1:N]}|\boldsymbol{\theta})$ is straightforward.
SBI offers a general solution to perform Bayesian inference in settings where the likelihood is either intractable or prohibitively expensive, but forward simulations are feasible. SBI starts by generating prior samples $\boldsymbol{\theta}_{i} \sim p(\boldsymbol{\theta})$, which are used to simulate synthetic trajectories $\boldsymbol{q}_{[1:N], i} \sim \mathcal{M}(\boldsymbol{\theta}_i)$ (Fig. \ref{fig:1} upper panel). The simulated data set $\mathcal{D} = \{(\boldsymbol{q}_{[1:N], i}, \boldsymbol{\theta}_{i})\}_{i=1}^{M}$ is then used to train a neural network-based density estimator $f_{\psi}(\boldsymbol{\theta}|\boldsymbol{q}_{[1:N]})$ of parameters $\psi$ to approximate the Bayesian posterior (Fig. \ref{fig:1} center panel). The neural network parameters $\psi$ are optimized by minimizing the loss function $\mathcal{L}(\psi) = 1/M\sum_i^M f_{\psi}(\boldsymbol{\theta}_{i} | \boldsymbol{q}_{[1:N], i})$ on $\mathcal{D}$. After training, the posterior for an experimental trajectory is obtained as $p(\boldsymbol{\theta}|\boldsymbol{q}_{[1:N]}^{\textrm{exp}}) \approx f_{\psi}(\boldsymbol{\theta}|\boldsymbol{q}_{[1:N]}^{\textrm{exp}})$, enabling inference of the folding parameters (Fig. \ref{fig:1} right panel). 

\begin{figure}[t!]
    \centering
    \includegraphics[width=0.45\textwidth]{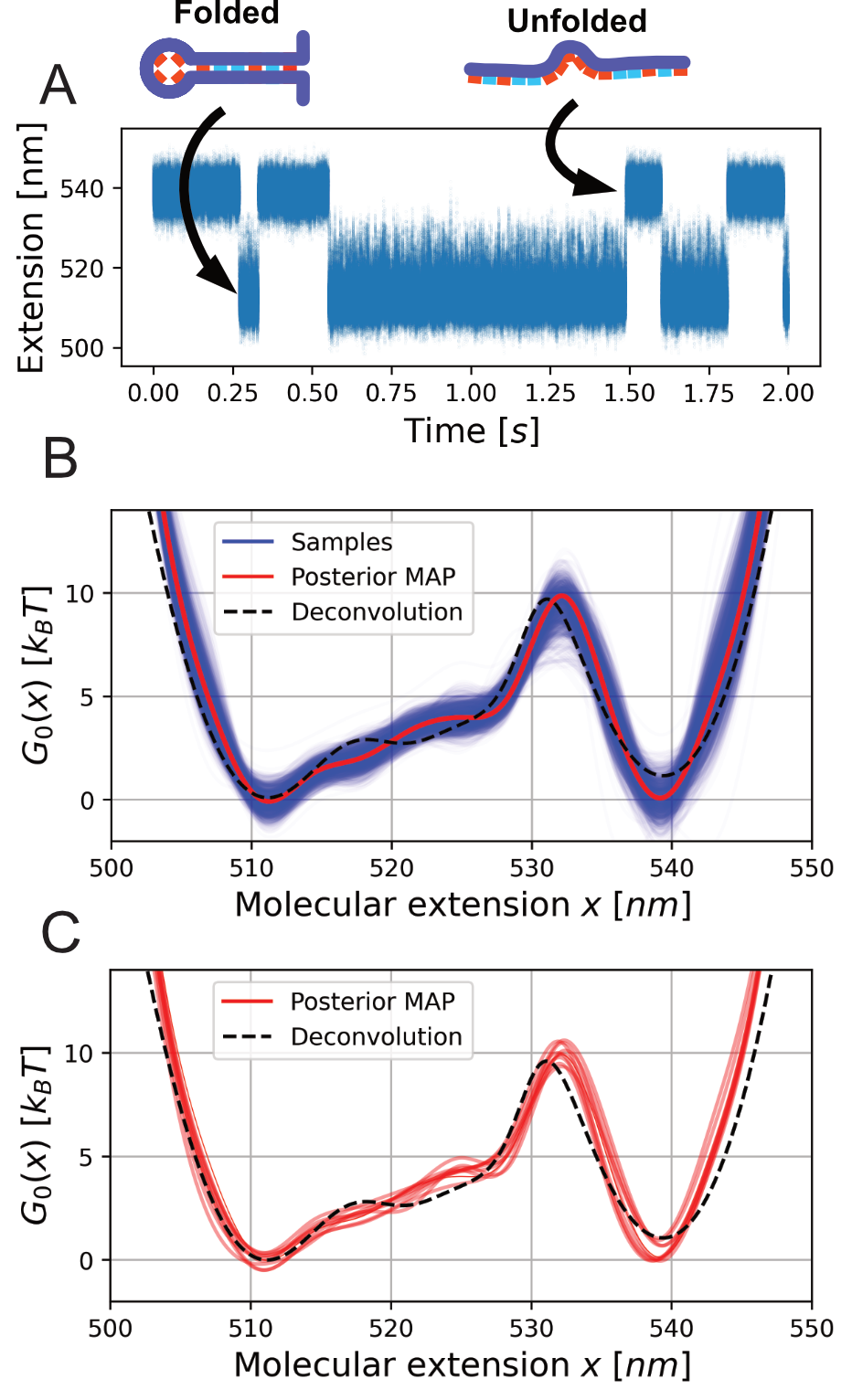}
    \caption{Free energy profile reconstruction. (A) Experimental time series used for inference. (B) Reconstructed free energy profile. Best estimate $\boldsymbol{\theta}^{\textrm{exp}}_{\textrm{MAP}}$ (Maximum a posteriori, MAP) in red, and posterior samples covering a 68 \% confidence interval as blue thin lines. 
    The black line indicates the estimate  using deconvolution. (C) Best free energy profile (MAP) estimate for 20 independent experimental time series.}
    \label{fig:2}
\end{figure}

We analyzed data from the 30R50/T4 DNA hairpin (Fig. \ref{fig:2}A Cartoon). Our analysis used a short 2-second long constant force measurement $\boldsymbol{q}_{[1:N]}^{\textrm{exp}}$ (Fig. \ref{fig:2} A) that captured approximately seven transitions between the folded and unfolded states.
We used Sequential Neural Posterior Estimation (SNPE) to approximate the posterior distribution $p(\boldsymbol{\theta}|\boldsymbol{q}_{[1:N]}^{\textrm{exp}})$ for the experimental trajectory $\boldsymbol{q}_{[1:N]}^{\textrm{exp}}$. The trained surrogate posterior can be evaluated on the experimental data, allowing parameters to be sampled from the posterior, $\boldsymbol{\theta}^{\textrm{exp}} \sim p(\boldsymbol{\theta}|\boldsymbol{q}_{[1:N]}^{\textrm{exp}})$, to explore many parameter combinations consistent with the experiment. Alternatively, identifying the maximum a posteriori estimate, $\boldsymbol{\theta}^{\textrm{exp}}_{\textrm{MAP}} = \textrm{argmax}_{\boldsymbol{\theta}}\ p(\boldsymbol{\theta}|\boldsymbol{q}_{[1:N]}^{\textrm{exp}})$, yields the best-fitting parameter set. SNPE approximates the posterior through iterative rounds of simulation and posterior density estimation. In each round, we carried out simulations with parameters drawn from the current posterior approximation. After 14 such rounds, each consisting of 2000 simulations, we obtained a surrogate posterior $p(\boldsymbol{\theta}|\boldsymbol{q}_{[1:N]}^{\textrm{exp}})$ (Fig. S1) in approximately 11 hours on a single Xeon Skylake Gold 6148 CPU.

Figure \ref{fig:2}A shows the experimental trajectory $\boldsymbol{q}_{[1:N]}^{\textrm{exp}}$ we analyzed. The folding free energy profile associated with the inferred model parameters $\boldsymbol{\theta}^{\textrm{exp}}$ is shown in Figure \ref{fig:2}B. The red line represents the best-fitting profile, obtained from the MAP parameters $\boldsymbol{\theta}^{\textrm{exp}}_{\textrm{MAP}}$. The profile shows a barrier height of approximately $9.9\ k_{\textrm{B}}T$. The thin blue lines show free energy profiles generated from posterior samples $\boldsymbol{\theta}^{\textrm{exp}}$ within the 68$\%$ confidence interval, illustrating the uncertainty of the estimate (see Fig. S2 for the 95\% confidence interval). Sampling parameters consistent with $\boldsymbol{q}_{[1:N]}^{\textrm{exp}}$ from the posterior distribution yield a median barrier height of $9.5 \pm 1.3\ k_{\textrm{B}}T$ (Fig. S3).

\begin{figure}[b!]
 \centering
 \includegraphics[width=0.45\textwidth]{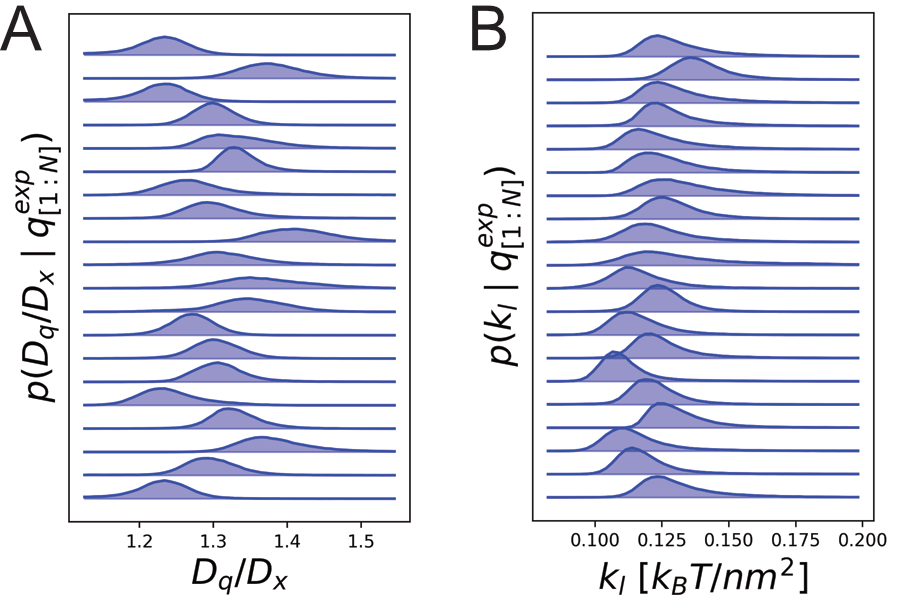}
 \caption{Diffusion coefficients and linker stiffness estimates.  Posteriors obtained using 20 independent experimental time series, quantifying the inference on (A) the ratio of diffusion coefficients $D_q / D_x$, and (B) the linker stiffness $k_l$. Posterior marginals were obtained by sampling and histogramming. All marginals are normalized to unit area. A constant vertical offset is applied between marginals to aid visualization}
 \label{fig:3}
\end{figure}

We validated our results by comparing our inferred free energy landscape with one obtained via deconvolution (Fig. \ref{fig:2}B black dotted line). Deconvolution is an established method that reconstructs the free energy landscape by iteratively removing measurement noise from the experimental data (see SI) \cite{woodside_direct_2006}. Using measurements of the biomolecule-linker system $G(q)$ and the linker only $V(q-x)$, deconvolution aims to find $G_0(x)$, that, when convolved with $V(q-x)$, reproduces $G(q)$. While both approaches yield closely matching profiles, confirming the reliability of our inference, deconvolution requires substantially more experimental data — approximately 20 - 100 times as many folding/unfolding transitions. This makes deconvolution more labor-intensive \cite{woodside_direct_2006} and prone to errors arising from instrumental drift. Furthermore, deconvolution requires a separate linker characterization, where inaccuracies can introduce errors in the resulting landscape.

\begin{figure}[t!]
 \centering
 \includegraphics[width=0.45\textwidth]{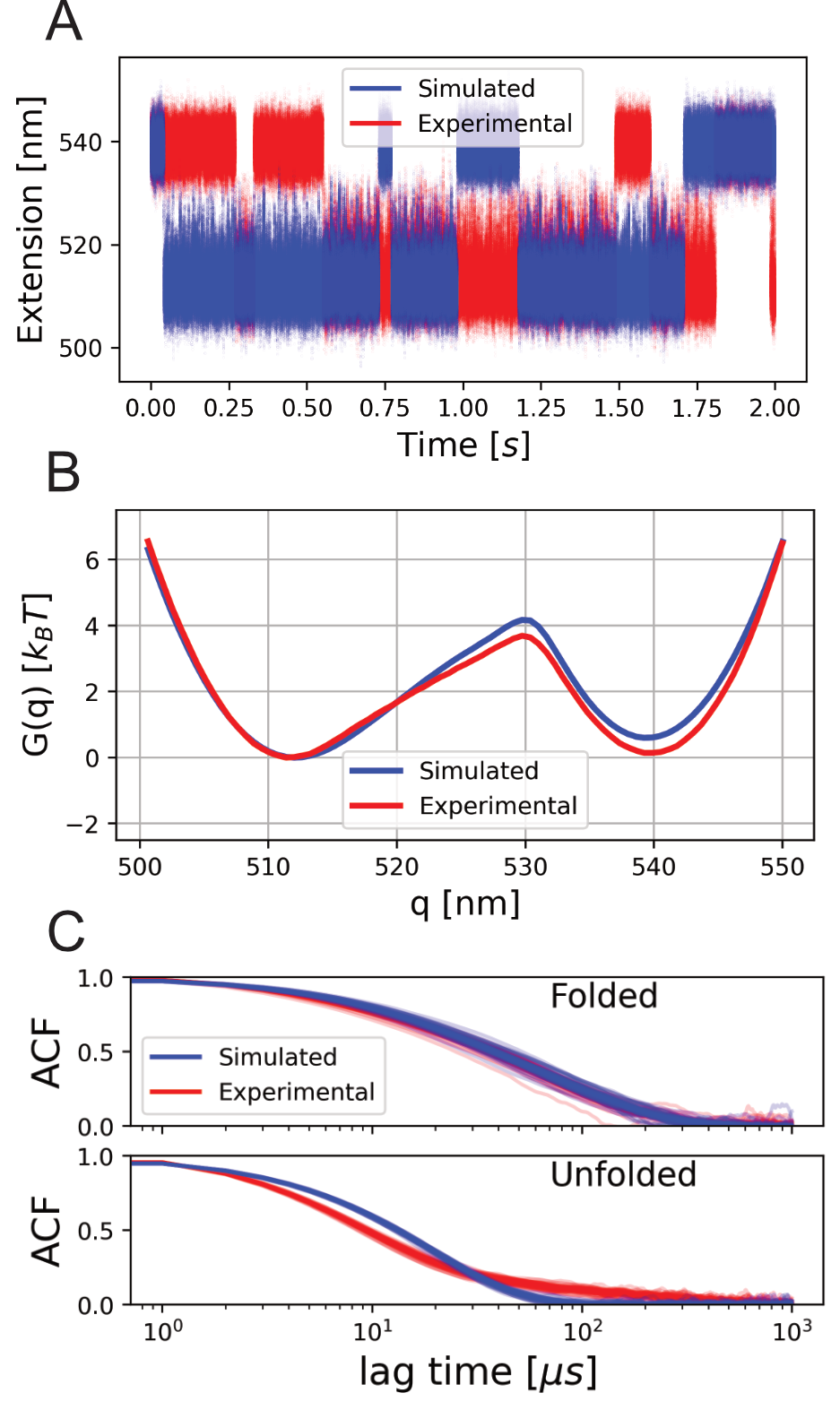}
 \caption{Predictive checks using simulations with best-fitting parameters $\boldsymbol{\theta}^{\textrm{exp}}_{\textrm{MAP}}$. (A) Trajectory $\boldsymbol{q}_{[1:N], 1}$ simulated with $\boldsymbol{\theta}^{\textrm{exp}}_{\textrm{MAP}, 1}$ (blue) compared to experimental trajectory $\boldsymbol{q}_{[1:N], 1}^{\mathrm{exp}}$ used to make the inference (red). (B) The potential of mean force estimated from 20 synthetic trajectories $\boldsymbol{q}_{[1:N], i}$ (blue) simulated with $\boldsymbol{\theta}^{\textrm{exp}}_{\textrm{MAP}, 1}$ compared to the potential of mean force obtained from the experimental trajectories (red). (C) Autocorrelation functions for segments of the trajectories in the folded and unfolded states, respectively, comparing experimental $\boldsymbol{q}_{[1:N],i}^{\mathrm{exp}}$ (red) and simulated trajectories $\boldsymbol{q}_{[1:N],i}$ (blue).}
 \label{fig:4}
\end{figure}

Additionally, we demonstrated that our estimated uncertainties are accurate by applying our inference framework to 20 independent 2-second measurements $\{\boldsymbol{q}_{[1:N],i}^{\mathrm{exp}}\}_{i=1}^{20}$ of the same DNA hairpin molecule. The results, shown in Figure \ref{fig:2}C, yield statistically consistent free energy landscapes across all measurements. Deviations among these independently inferred profiles (red lines in Figure \ref{fig:2}C) fall within the uncertainty bounds (blue lines in Figure \ref{fig:2}B) determined from analyzing a single trajectory, indicating that the error model of the surrogate posterior is well-calibrated.

The model we fitted to the experimental data $\boldsymbol{q}_{[1:N]}^{\textrm{exp}}$ also obtained estimates for the ratio of diffusion coefficients $D_q/D_x$ and linker stiffness $k_l$. Figure \ref{fig:3} reports the marginal posterior distribution for  $D_q/D_x$ and $k_l$ for 20 independent experimental trajectories. We also observed a satisfactory overlap of the marginal posterior distributions for these parameters, although larger than expected, pointing to the existence of model misspecification, in particular for the diffusion coefficients (see SI).

To further assess our model's accuracy and predictive power, we simulated trajectories using the best-fitting parameter, i.e., $\boldsymbol{q}_{[1:N], i} \sim p(\boldsymbol{q}_{[1:N]} \mid \boldsymbol{\theta}^{\textrm{exp}}_{\textrm{MAP}, 1})$. We then compared these simulated trajectories to the experimental trajectory used for parameter inference (Fig. \ref{fig:4}A). If the model is adequate and the fit accurate, we expect that the experimental $\boldsymbol{q}_{[1:N]}^{\mathrm{exp}}$ and simulated trajectory $\boldsymbol{q}_{[1:N]}$ look very similar. 

Visual comparison shows excellent agreement between the simulated and experimental trajectories. To quantify this, we performed a total of 20 simulations $\{\boldsymbol{q}_{[1:N],i}\}_{i=1}^{20}$ using the best fitting parameter $\boldsymbol{\theta}^{\textrm{exp}}_{\textrm{MAP}, 1}$. First, we compared the thermodynamics of the observed trajectories by inspecting the potential of mean force (PMF) $G(q)$. We computed the average PMF $G(q)$ by combining all trajectories and binning them along $q$.  The simulated and experimental PMFs align well in the folded and unfolded states, though the simulated trajectories exhibit a slightly higher barrier in the transition region (Fig. \ref{fig:4}B). The variability of the individual PMFs is shown in Fig. S4.

We also compared the observed kinetics by analyzing folding and unfolding transitions (see SI). The experimental transition rate of $2.8\pm 0.3 ~\mathrm{s^{-1}}$ matches the simulated rate of $2.2\pm 0.2~\mathrm{s^{-1}}$, demonstrating that our model captures the underlying kinetics. This agreement between the rates demonstrates that our model successfully captures the underlying kinetics.

Finally, we compared the autocorrelation functions (ACF) of the measured extension in the folded and unfolded states (Fig. \ref{fig:4}C). While the ACFs align well in the folded state (upper panel), the experimental trajectories in the unfolded state exhibit a non-single exponential decay not captured by our simulations (lower panel). This discrepancy is consistent with previously reported memory effects for the 30R50/T4 hairpin \cite{pyo_memory_2019}. This suggests that a Markovian diffusion model may not fully describe the complex dynamics of an unfolded DNA hairpin, possibly because the dynamics  are not 1D \cite{satija_transition_2017, satija_generalized_2019, pierse_distinguishing_2017, satija_broad_2020}.

\begin{figure}[t!]
 \centering
 \includegraphics[width=0.45\textwidth]{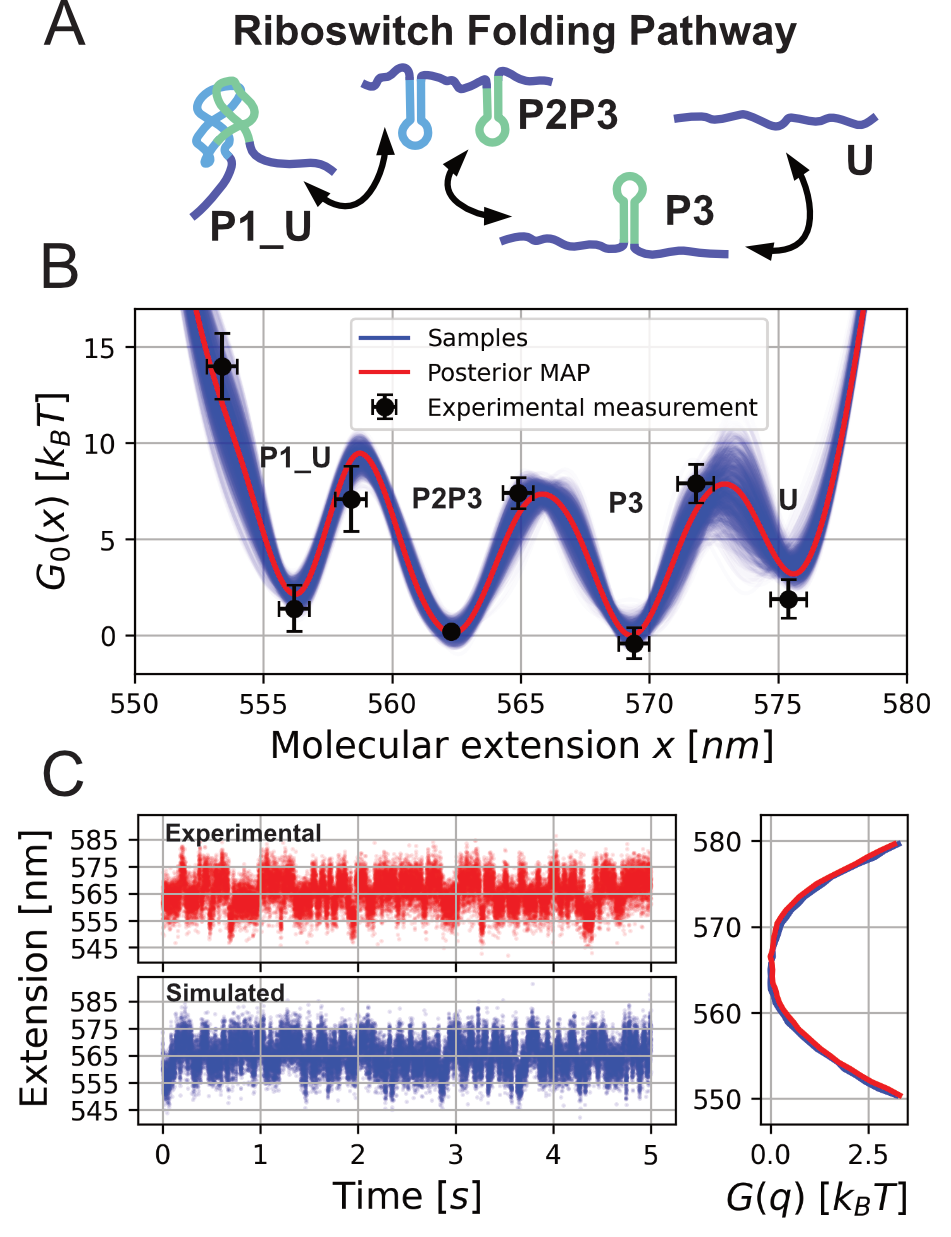}
 \caption{Riboswitch folding. (A) Schematic illustration of the folding pathway observed during the constant force experiment. Beginning from the unfolded state U, two hairpins, P2 and P3, are formed sequentially and subsequently establish a tertiary contact, P1\_U. (B) Reconstructed free energy profile. Best estimate $\boldsymbol{\theta}^{\textrm{exp}}_{\textrm{MAP}}$ shown in red, and 68 \% confidence interval shown in blue. Positions and energies of the potential wells and barriers (black) previously deduced from experiments \cite{neupane2011single}, measured relative to state P2P3. (C) Comparison between the experimental trajectory (red) and a simulated trajectory (blue) generated using $\boldsymbol{\theta}^{\textrm{exp}}_{\textrm{MAP}}$. Potential of mean force computed from the experimental (red) and simulated trajectory (blue).}
 \label{fig:5}
\end{figure}

To show that the same analysis framework can be applied to molecules with more complex folding, including multiple intermediate states and formation of tertiary contacts, we also analyzed the folding of an RNA riboswitch apatmer. These non-coding mRNA elements regulate gene expression via ligand-dependent structural transitions \cite{roth2009structural}. We analyzed a five-second-long constant force measurement $\boldsymbol{q}^{\mathrm{exp}}_{[1:N]}$ of the \textit{add} riboswitch aptamer, whose folding pathway involves five states, where in this particular trajectory four out of five states are explored \cite{neupane2011single}. The riboswitch folds by sequentially forming two hairpins that then form a ligand-binding pocket organized by tertiary contacts (Fig. \ref{fig:5}A). We used SNPE on $\boldsymbol{q}^{\mathrm{exp}}_{[1:N]}$ to infer a folding model of the riboswitch (Supplementary Material).
The folding free energy landscape shows four distinct states, each corresponding to one of the folding intermediates we expected to observe (Fig. \ref{fig:5}B). The landscape is consistent with results from previous single-molecule studies \cite{neupane2011single}, identifying positions and energies of the potential wells and barriers that match previous analyses within error (Figure 5B, black). Finally, predictive checks validate the fitted model, as trajectories $\boldsymbol{q}_{[1:N]}$ simulated with $\boldsymbol{\theta}^{\textrm{exp}}_{\textrm{MAP}}$ are consistent with the experimental measurement $\boldsymbol{q}_{[1:N]}^{\textrm{exp}}$. The potentials of mean force $G(q)$, computed from the simulated (blue) and experimental trajectory (red) also show excellent agreement (Fig.~\ref{fig:5}C)\\\\

Characterizing the microscopic dynamics in biomolecular folding remains a significant challenge in biophysics. Single-molecule force spectroscopy provides direct insights into folding mechanics, though instrumental noise and artifacts complicate the extraction of properties like energy landscapes. To address these issues, we have developed a framework based on simulation-based inference that enables us to learn quantitative folding models from limited experimental data.

Using a DNA hairpin measured at constant force, we learned a quantitative folding model that describes the observed data. This approach estimates free energy landscapes that match the results from established deconvolution methods while requiring significantly less data and eliminating the need for separate control measurements of instrumental artifacts. Our method also provides estimates of diffusion coefficients and linker properties. The Bayesian framework offers a key advantage: it generates complete posterior distributions that go beyond point estimates, allowing us to quantify uncertainties with testable reliability, while requiring much less data.

We validated the learned models by comparing experimental trajectories with those simulated using the best-fitting parameters, finding that our model reproduces the experimental data in a quantitatively robust and predictive manner. Minor discrepancies between the model and experimental data remain. Additional analysis using trajectories of different length (Fig.~S5) suggests that these errors are unlikely to arise from having insufficient data---indeed, even $1$-s trajectories of the hairpin allowed high-quality inference of the landscape. Going further, reliable inference does not even require single long trajectories: short, independent segments are enough to recover the same landscape. For the DNA Hairpin, this means that with as few as eight $0.125$ s segments ($1$~s total) we obtained results comparable to those obtained from full $2$~s long trajectories (Fig.~S11). This makes the framework directly applicable to high-throughput parallel SMFS experiments, where many short trajectories from many molecules are the natural data output rather than a few long ones.

The residual discrepancies between model and data therefore likely arise from other sources, such as non-Markovian dynamics or position-dependent diffusion in the experimental data \cite{pyo_memory_2019, hoffer2021observing}. Such deviations suggest the need for a more complex model to capture the underlying folding dynamics more fully. Despite the simplicity of the model used in our analysis, our approach was still successful when extended from a simple two-state hairpin to a more complex structure featuring multiple intermediates and tertiary contacts.

To improve the predictive capabilities of the framework, two approaches could potentially be used. First, we could incorporate memory effects by utilizing a more complex dynamical model, which would improve predictiveness but introduce parameters that lack direct physical interpretation. Alternatively, we could use molecular dynamics simulations as the simulator $\mathcal{M}(\theta)$, allowing us to infer molecular properties from single-molecule trajectories. While this approach offers the highest fidelity, it does so at a higher computational cost. In general, while more complex dynamical models can be integrated into the simulation-based inference framework, they introduce additional methodological and computational challenges that remain to be addressed in future work.

Extracting statistically reliable results from limited datasets is important for investigating complex systems where extensive data collection is impractical. By fitting a dynamical model directly to observed data, SBI leverages temporal correlations, extracting more information than methods that ignore these relationships. Our method efficiently constructs comprehensive folding models for diverse biomolecular systems. It extends naturally to other single-molecule force spectroscopy protocols, such as constant trap experiments, offering a unified framework for various experimental datasets.\\\\

\noindent
\textit{Acknowledgement - }We thank Dr Attila Szabo for useful discussions and feedback. L.D. and R.C. acknowledge the support of Goethe University Frankfurt, the Frankfurt Institute of Advanced Studies, the LOEWE Center for Multiscale Modelling in Life Sciences of the state of Hesse, the CRC 1507: Membrane-associated Protein Assemblies, Machineries, and Supercomplexes (P09), and the International Max Planck Research School on Cellular Biophysics, as well as computational resources and support from the Center for Scientific Computing of the Goethe University and the Jülich Supercomputing Centre. The Flatiron Institute is a division of the Simons Foundation. This work was supported by Natural Sciences and Engineering Research Council of Canada (grant reference number RGPIN-2018-04673, to MTW).\\\\

\noindent
\textit{Data availability - } The code is available at GitHub \url{https://github.com/covinolab/SBIsmfs} and is based on the SBI-toolkit \cite{boelts_sbi_2024}, a PyTorch-based implementation of simulation-based inference algorithms.
Data and scripts necessary to reproduce all the results presented in this paper are freely accessible at the Zenodo repository \url{https://zenodo.org/record/19881932}.\\\\

\newpage
\onecolumngrid
\thispagestyle{empty}
\begin{center}
\vspace*{\fill}
{\Large\bfseries Supplementary Material}
\vspace*{15pt}
\end{center}
\twocolumngrid

\section{Simulator}
We used the harmonic-linker model, which is well-established for single-molecule force spectroscopy experiments \cite{cossio_artifacts_2015, covino_2019, hummer_free_2010}.
This model describes the coupled system of biomolecule and apparatus by Brownian diffusion on a 2-dimensional free energy surface. Thereby, the (hidden) molecular extension $x$ is defined by the distance between the two attachment points on the biomolecule. The measured extension $q$ describes the experimental observable, the distance between the ends of the linker.
The free energy surface $G(q, x)$ describes the coupled system of molecule and apparatus:
\begin{equation}
    G(x, q) = G_{0}(x) + \frac{k_l}{2}(x - q)^2
\end{equation}
Here, $G_0(x)$ describes the molecular free energy surface, including the constant pulling force, which is applied during constant-force single-molecule experiments. We modeled the free energy surface $G_{0}(x)$ using cubic spline interpolation. The parameter $k_l$ describes the stiffness of the harmonic linker, where $(x-q)$ is the linker extension. We assume an anisotropic $(D_x \neq D_q)$ and position-independent diffusion coefficients. Here, $D_x$ describes the intrinsic diffusion coefficient of the molecule, and $D_q$ describes the diffusion coefficient of the pulling apparatus. Because $D_x$ is degenerate with the barrier heights of $G_0(x)$, only the ratio of $D_q/D_x$ is identifiable. By simulating Brownian dynamics on the free energy surface $G(q, x)$, we obtain trajectories of the molecular extension $x(t)$ and the corresponding observed extension $q(t)$.

\subsection{Brownian simulations on $G(q, x)$}
The simulator $\mathcal{M}(\boldsymbol{\theta)}$ simulates the system according to Brownian dynamics. We used the Euler-Maruyama scheme to simulate the dynamics.
\begin{align}
    q(t + \Delta t) &= q(t) - \Delta q \\
    \Delta q &= \beta \partial_{q}G(q, x)\cdot D_q\Delta t - \sqrt{2D_q\Delta t}\cdot R_q(t) \\
    x(t + \Delta t) &= x(t) - \Delta x \\
    \Delta x &=  \beta \partial_{x}G(q, x)\cdot D_x\Delta t - \sqrt{2D_x\Delta t}\cdot R_x(t)
\end{align}
Here, $D_{x}$ and $D_{q}$ are the diffusion coefficients along $q$ and $x$. $R_q(t)$ and $R_x(t)$ are uncorrelated Gaussian noise with zero mean and unit variance. $\Delta t$ is the integration time step. $M_{\textrm{int}}$ is the total number of integration steps, so the total simulated time is $T = M_{\textrm{int}}\Delta t$. Each simulated trajectory is sub-sampled to match the saving frequency of the observed trajectory $\boldsymbol{q}_{[1:N]}^{\textrm{exp}}$ by a factor of $\Delta\nu$ so that $\boldsymbol{q}_{[1:N]}
= \{q(\Delta t \Delta\nu \cdot i)\}_{i=1}^{M_{\textrm{int} / \Delta\nu}}$.

\section{Modeling $G_{0}(x)$}
The molecular free energy surface $G_{0}(x)$ is modeled using cubic spline interpolation. We used $N_{\textrm{node}}$ equally separated spline nodes at positions $x_i$. Each node has one adjustable parameter $G_{0}(x_i)$ that determines its height, with more nodes enabling a finer description of the energy surface. To model the space between adjacent nodes, we used cubic spline functions of the form $C_i(x) = a_i + b_i x + c_i x^2 + d_i x^3$. For $N_{\textrm{node}}$ nodes, we constructed $N_{\textrm{node}}-1$ spline functions, where $C_i(x)$ describes the free energy in the interval between $x_i$ and $x_{i+1}$. For a given set of $\{G_{0}(x_i)\}_{i=1}^{N_{\textrm{node}}}$, the spline parameters $\{(a_i$, $b_i$, $c_i$, $d_i)\}_{i=1}^{N_{\textrm{node}}}$ are fitted to ensure continuous first and second derivatives at the node junctions. Once these parameters are determined, we can efficiently evaluate both the spline function and its derivatives. To confine the system within the defined energy landscape, we introduced a positive energy offset at the boundary nodes. The first two and last two nodes were shifted upwards relative to their neighboring nodes. The free energy profiles are defined up to an additive constant.

\section{Neural Posterior Estimation}
We employed Sequential Neural Posterior Estimation (SNPE) \cite{lueckmann_benchmarking_2021} to learn a surrogate model of the Bayesian posterior $p(\boldsymbol{\theta} | \boldsymbol{q}_{[1:N]})$. The surrogate posterior combines two neural networks: a density estimator $f_{\psi}(\boldsymbol{\theta} | S_{\Psi}(\boldsymbol{q}_{[1:N]}))$ of parameters $\psi$ to estimate posterior density over model parameters $\boldsymbol{\theta}$, and an embedding network $S_{\Psi}(\boldsymbol{q}_{[1:N]})$ of parameters $\Psi$ that learned lower-dimensional summary statistics from the input. Both networks were trained jointly using the same loss function.
For the density estimator, we used a normalizing flow (NF), which consists of a series of invertible mappings between a base Gaussian distribution and the target posterior distribution. These mappings were parameterized by neural networks.
In our case, we used the Neural Spline Flow \cite{durkan_neural_2019} and Masked Autoregressive Flows \cite{papamakarios_masked_2017} as the choice for the normalizing flows. The embedding network $S_{\Psi}(\boldsymbol{q}_{[1:N]})$ consisted of two parts.  First, the input time series were featurized using transition matrices (see Timeseries Featurization), which were then used as the input into a single-layer convolutional neural network. The joint network is trained on a simulated dataset $\mathcal{D} =\{ \boldsymbol{\theta}_{i}, \boldsymbol{q}_{[1:N], i})_{i=1}^M$ with $M$ samples, generated by drawing parameters $\boldsymbol{\theta}_i \sim p(\boldsymbol{\theta})$ from the prior and using them to run simulations with the forward model $\boldsymbol{q}_{[1:N], i} \sim p(\boldsymbol{q}_{[1:N]}|\boldsymbol{\theta}_i)$. The surrogate posterior was trained by maximizing the average log-likelihood:
\begin{align}
\label{eq: amortized_loss}
\mathcal{L}(\psi, \Psi) = \sum_i^M log(f_{\psi}(\boldsymbol{\theta}_{i} | S_{\Psi}(\boldsymbol{q}_{[1:N], i}))
\end{align}
After training, the surrogate model approximates the true posterior, $f_{\psi}(\boldsymbol{\theta} | S_{\Psi}(\boldsymbol{q}_{[1:N]})) \approx p(\boldsymbol{\theta}|\boldsymbol{q}_{[1:N]})$.
In our case, we used SNPE \cite{greenberg_automatic_nodate}, as it requires fewer simulations to approximate the posterior for a single observation. SNPE is not amortized. Therefore, a new surrogate posterior must be retrained for each new observation, offering greater simulation efficiency for individual inferences.
The SNPE inference process consists of $K$ rounds, in each round $K_{\text{sim}}$ simulations are generated and added to the data set $\mathcal{D}$. Afterwards, the surrogate posterior is trained on the extended dataset. In the first round, parameters are drawn from the prior distribution $\boldsymbol{\theta}_i \sim p(\boldsymbol{\theta})$, and the surrogate posterior is trained using the loss function in Equation \ref{eq: amortized_loss}. In subsequent rounds, parameters are generated from the previously trained surrogate posterior $\boldsymbol{\theta}_i \sim f_{\psi}(\boldsymbol{\theta} | S_{\Psi}(\boldsymbol{q}_{[1:N]}))$ and used to generate new simulations $\boldsymbol{q}_{[1:N], i} \sim p(\boldsymbol{q}_{[1:N]}| \boldsymbol{\theta}_i)$. These new simulations, which should be more similar to the observed data $\boldsymbol{q}_{[1:N]}$, are added to the dataset $\mathcal{D}$.
Specifically, we use the SNPE(C) \cite{greenberg_automatic_nodate} implementation from the SBI-toolkit \cite{boelts_sbi_2024} to correct for prior changes when training the density estimator over multiple rounds. After $K$ rounds of simulation and training, the final surrogate posterior approximates the true posterior: $f_{\psi}(\boldsymbol{\theta} | S_{\Psi}(\boldsymbol{q}_{[1:N]})) \approx p(\boldsymbol{\theta}| \boldsymbol{q}_{[1:N]})$.

\section{Timeseries Featurization}\label{sec: time series}
We compressed the measured trajectories $\boldsymbol{q}_{[1:N]}$ into a medium-dimensional vector containing summary statistics. As summary statistics, we used the elements of the estimated transition matrices  $T_{ik}(\Delta\tau)$ at different lag times $\Delta\tau$. To estimate each transition matrix, we segmented the trajectory $\boldsymbol{q}_{[1:N]}$ into $n_{\textrm{bins}}$ equally spaced bins. The matrix elements were populated by counting transitions between bins $i$ and $k$ at each lag time, followed by column-wise normalization. This procedure is repeated for multiple lag times to capture the system's dynamics across different timescales. To concatenate trajectories and perform inference on multiple segments, the transition counts from each trajectory were added together and then normalized.

\section{Priors}
We factorized the prior $p(\boldsymbol{\theta})$ over the parameters $\boldsymbol{\theta} = \{D_q/D_x, k_l, G(x_{0}), ..., G(x_{N_{\textrm{node}}})\}$ into the individual components: $p(\boldsymbol{\theta}) = p(D_q/D_x)\cdot p(k_l)\cdot \prod_i^{N_{\textrm{node}}} p(G_0(x_i))$. The prior for each individual parameter is either modeled as a Gaussian distribution $\mathcal{N}(\mu, \sigma^2)$ with mean value $\mu$ and variance $\sigma$, or a uniform distribution $\mathcal{U}(a, b)$ with lower bound $a$ and upper bound $b$. For the ratio of diffusion coefficients $D_q/D_x$ and linker stiffness $k_l$, we used log-uniform distributions. For the spline nodes, we used a multivariate Gaussian model in which the heights of the spline nodes are correlated. This allowed us to encode a rough shape of the free energy profile into the prior. The mean of the multivariate Gaussian $\boldsymbol{\mu}$ encodes the most likely a priori known free-energy profile. The covariance matrix of the multivariate Gaussian $\boldsymbol{\sigma}_{i, j}$ encodes how correlated adjacent nodes are, which allows us to enforce smoothness. The covariance matrix was constructed with the following formula:
\begin{align*}
    \boldsymbol{\sigma}_{i, j} = \alpha \cdot e^{\beta |i - j|}
\end{align*}
With $i$ and $j$ going from $0$ to $N_{\textrm{node}}$ the number of adjustable nodes. The parameter $\alpha$ scales the covariance matrix and determines the overall width of the Gaussian, while the parameter $\beta$ determines the correlation between adjacent nodes.
In principle, prior predictive checks can be used as a diagnostic tool to assess whether the chosen priors lead to unrealistic model behavior and to identify poorly specified priors.

\section{Rate Estimation}
We determined the transition rate $k$ by counting the total number of transitions $n_t$ (both folding and unfolding events) observed in the measured trajectories. To estimate $k$, we maximized the Poisson likelihood function $p(n_t\mid t,k)$, where $t$ represents the trajectory duration.\begin{align*}
    p(n_{t}|t, k)~=~\frac{(kt)^{n_{t}}\cdot e^{-kt}}{n_{t}!}
\end{align*}
The maximum likelihood estimate for the Poisson distribution yields $k_{\mathrm{max}}=n_{t}/t$. To quantify the uncertainty in our rate estimate, we calculated the 68\% confidence interval by integrating the Poisson likelihood function. We identified the $k$ values where the integrated and normalized likelihood encompasses 16\% and 84\% of the probability density. This is equivalent to a Bayesian error estimation with a uniform prior distribution.

\section{Hyperparameters}
\subsection{30R50T4 DNA hairpin}
The simulations used a time step of $\Delta t = 0.01\ \mu s$, with a total of $M_{\textrm{int}} = 2,000,000$ integration steps. We saved the trajectories every $\Delta\nu = 100$ steps to match the experimental measurement frequency. The molecular diffusion coefficient was set to $D_x = 0.397\ nm^2/ \mu s^2$. The estimate was derived by following the approach described by Cossio et al. (2015)\cite{cossio_artifacts_2015} while averaging over all trajectories. The free energy profile has a total of 15 nodes, while the first and last nodes are always offset by 30 $k_{\textrm{B}}T$ and the second and second last nodes are always offset by 15 $k_{\textrm{B}}T$. Therefore, the model has $N_{\textrm{node}}=11$ adjustable nodes. The prior for the ratio of diffusion coefficients $D_q/D_x$ was a log-uniform distribution, with the lower and upper limits at $10^{-1}$ and $10^1$. The prior for the linker stiffness $k_l$ was also a log-uniform distribution with the limits set to $10^{-2}\ k_{\textrm{B}}T/nm^2$ and $10^0\ k_{\textrm{B}}T/nm^2$. The prior for the spline nodes was modeled with a multivariate Gaussian. The mean value was set to the average PMF $G(q)$ of the experimental trajectories. The parameters that define the amplitude $\alpha$ and correlation $\beta$ of the covariance matrix were set to 20 and 0.2, respectively. 
The trajectories were binned into $n_{\textrm{bins}} = 20$ equally sized bins between 510 and 540 $nm$. The transition matrices were computed for lag times $\Delta\tau$ of 1, 10, 100, 1000, 50000, and 500000 steps.
For the neural network, we used a neural spline flow (NSF) as the density estimator and a single convolutional neural network layer for the embedding network. We used the NSF implementation available in the SBI-toolkit\cite{boelts_sbi_2024}. The NSF consisted of five transformation stages, each parametrized by one neural network. The neural networks consisted of two blocks, each with 100 hidden features and 10 bins. We used a batch size of 64 with a learning rate of 0.00025. If the validation loss did not improve within 15 epochs, the training was terminated. We did not require extensive hyperparameter tuning, our choices follow standard practices and established benchmarks \cite{lueckmann_benchmarking_2021} in the simulation-based inference literature. 
We selected the posterior obtained after 14 training rounds for our final analysis since we observed that beyond this point, the 99$\%$ confidence intervals of posteriors from different trajectories began not to overlap anymore (Fig. \ref{fig:SI6}). This behavior indicates posterior miscalibration in later sequential rounds. We therefore restrict the number of rounds based on posterior consistency and validation-based early stopping during training phases.\\

We verified that the inferred posteriors are robust to numerical and modeling choices by performing additional analyses that varied trajectory length, numerical time discretization, and spline-based free energy parameterization. Specifically, we compared inference results obtained from analyzing a single 1-s trajectory to those from two 2-s trajectories (see Time series featurization; Fig. \ref{fig:SI5}), evaluated the impact of the numerical integration step by reducing $\Delta t$ by factors of 0.5 and 0.1 using a single 2~s trajectory (Fig. \ref{fig:SI7}, and assessed sensitivity to the spline representation by varying the number of spline knots (10 and 20), and altered the height of the boundaries (Fig. \ref{fig:SI8}). In all cases, all other hyperparameters were kept identical to the original analysis, and no substantial changes in the inferred posteriors were observed.

\subsection{Riboswitch}
Simulations were performed with a time step of $\Delta t = 0.01\ \mu s$ over a total of $M_{\textrm{int}} = 5\times10^{8}$ integration steps. Trajectories were saved every $\Delta\nu = 5000$ steps to match the experimental sampling rate. The molecular diffusion coefficient was set to $D_x = 0.116\ \mathrm{nm}^2/\mu s^2$ following the same routine as before.
The free energy profile was represented using 15 spline nodes, with the first and last nodes fixed at an offset of $30~k_{\textrm{B}}T$ and the second and second-last nodes fixed at $15~k_{\textrm{B}}T$, leaving $N_{\textrm{node}} = 11$ adjustable nodes. The prior on the diffusion coefficient ratio $D_q/D_x$ and the linker stiffness $k_l$ were chosen as log-uniform distributions over $[10^{-1}, 10^{1}]$ and $[10^{-2}, 10^{0}]\ k_{\textrm{B}}T/\mathrm{nm}^2$, respectively. The spline node values were assigned a multivariate Gaussian prior with a mean given by the average experimental PMF $G(q)$. To estimate the PMF, we first applied a moving average with a window of 50 data points to smooth the trajectory, and then binned the resulting smoothed data. This approach provided an approximate shape of the free energy profile. The covariance amplitude and correlation parameters were set to $\alpha = 30$ and $\beta = 0.1$.
Trajectories were discretized into $n_{\textrm{bins}} = 50$ equally spaced bins between 552 and 578 nm, and transition matrices were computed at lag times $\Delta\tau = 1, 10, 100, 1000,$ and $20000$ steps.
For inference, we employed a masked autoregressive flow (MAF) as the density estimator with a single convolutional layer as the embedding network, using the implementation provided in the SBI toolkit \cite{boelts_sbi_2024}. The MAF consisted of six transformation stages, each parameterized by a neural network with three blocks of 64 hidden units. Training was performed with a batch size of 64 and a learning rate of $2.5\times10^{-4}$, with early stopping triggered after 15 epochs without validation loss improvement. The posterior obtained after 10 training rounds of 3000 simulations each was selected for analysis, as it yielded the best posterior predictive agreement.

\section{Generalized Jensen–Shannon Divergence}
To qualitatively measure the similarity between $N$ different distributions $\boldsymbol{{p}^{i}}$, we computed the Generalized Jensen-Shannon Divergence (GJSD), which is defined as the difference between the Shannon entropy H of the average distribution and the average of the individual entropies \cite{nielsen2020generalization, englesson2021generalized}:
\begin{equation}
    D_{\text{GJS}}(\boldsymbol{p}^{1}, \dots, \boldsymbol{p}^{N}) = H\left(\frac{1}{N}\sum_{i=1}^N \boldsymbol{p}^{i}\right) - \frac{1}{N}\sum_{i=1}^N H(\boldsymbol{p}^{i}),
\end{equation}
with $H(p) = \mathbb{E}[-\log(p)]$.
We calculated the GJSD for the 20 posterior distributions and for their marginals regarding $D_q/D_x$ and $k_l$. For the full posterior distributions, we measured 1.3, while for the marginals of $D_q/D_x$ and $k_l$, we obtained values of 0.45 and 0.20, respectively. 

To establish a baseline for the expected scale of this overlap, we compared these results to a synthetic benchmark. For this purpose, we used the posterior estimate from one of the time series and used its MAP estimate to perform 20 independent simulations. We then computed the posterior for these 20 trajectories and determined their GJSD. For the full posterior distributions of the synthetic benchmark, we obtained a GJSD of 0.14, while the marginals (Fig. \ref{fig:SI9}) for $D_x/D_q$ and $k_l$ yielded 0.04 and 0.03.

The GJSD was computed using a Monte Carlo estimator where each distribution provides an equal contribution, following the expression\cite{nielsen2020generalization, englesson2021generalized}:
\begin{equation}
    D_{\text{GJS}} \approx \frac{1}{N} \sum_{i=1}^N \left[ \frac{1}{S} \sum_{k=1}^S \log \left( \frac{\boldsymbol{p}^i(x_{i,k})}{\frac{1}{N} \sum_{j=1}^N \boldsymbol{p}^j(x_{i,k})} \right) \right]
\end{equation}
where $N$ is the number of distributions and $S$ is the number of samples $x_{i, k}$ drawn from the distribution $\boldsymbol{p}^{i}$. For the full posterior, we utilized 100,000 samples. To estimate the GJSD for the marginals, we first estimated the density using a Gaussian Kernel Density Estimate (KDE) and subsequently used this to calculate the divergence. We utilized 10,000 samples per marginal to estimate the GJSD.

\section{Hairpin Analysis on Cropped Trajectory Segments}\label{sec: cropped}

To assess the data requirements of our framework, we performed inference on cropped segments of the experimental 30R50/T4 hairpin trajectories. We considered both inference on single short segments, to test how short an individual trajectory can be while still allowing reliable inference, and inference on multiple independent short segments. This was motivated by the idea of high-throughput parallel SMFS experiments in which the natural data product consists of many short trajectories rather than a few long ones.

\subsection{Cropping}
From the $20$ experimental trajectories, we randomly cropped segments of length $L \in \{0.125, 0.25, 0.5\}$~s. We then grouped them into datasets containing $N \in \{1, 2, 4, 8, 16\}$ segments each. For each $(N, L)$ combination, we required the total duration of the dataset to be smaller than or equal to the duration of the original trajectories (2s).

For each $(N, L)$ combination, segments were drawn at random from the pool of $20$ trajectories. To assess the variability of the inference results across different random selections, we generated $8$ independent replicates per $(N, L)$ combination, each constructed from a fresh random draw of segments. Representative examples of the resulting cropped datasets are shown in Fig.~SI10.

\subsection{Inference}
Inference was performed independently on each replicate using the same SBI pipeline, priors, and network architecture as in the main hairpin analysis (see Hyperparameters, 30R50T4 DNA hairpin). Two adjustments were made to accommodate the shorter segment lengths.

First, the set of lag times used to construct the transition matrices that summarize the trajectories (see Section~\ref{sec: time series}) was adjusted so that the maximum lag time did not exceed the segment length $L$. 

Second, when a dataset consisted of multiple segments ($N > 1$), the transition counts obtained from individual segments were added, as described in Section~IV.

\subsection{Results}
Figure~\ref{fig:SI11} shows the free-energy landscapes obtained from the MAP estimates for all $(N, L)$ combinations and replicates, alongside the $20$ reference landscapes inferred from the full $2$~s trajectories of the main analysis. Single short segments yield landscapes that vary substantially across replicates and only partially match the reference, since such segments typically contain zero or one transition and are therefore not guaranteed to be representative of the underlying dynamics. Combining a small number of independent short segments restores reliable inference: as few as eight $0.125$~s segments ($1$~s total) yield results comparable to those obtained from the full $2$~s trajectories. The framework therefore, does not require a single long trajectory, but it requires only that the combined data sample the relevant transitions sufficiently often.

These results probably depend on the system, the sampling rate, and the intrinsic kinetics of the molecule. We recommend that users perform an analogous test on synthetic data generated for their own system before applying the method to experimental data, in order to identify the regime in which reliable inference can be expected.

\section{Experimental Setup}
\subsection{30R50T4 DNA hairpin}
The DNA hairpin data presented in this work were first presented in Ref. \cite{lyons_quantifying_2024}. Briefly, the hairpin was attached to kilobase-length DNA handles containing either a biotin or digoxigenin molecule at their 5’-end using autosticky polymerase chain reaction (PCR). This construct was then attached to sub-micron diameter beads coated with either avdin or anti-digoxigenin, allowing the DNA hairpin to be held between two independently controlled optical traps. Tension could then be applied to the hairpin by increasing the distance between the two optical traps, which were separated until the hairpin began to spontaneously unfold. Once the hairpin was observed to occupy its folded and unfolded states with approximately equal probability (at a force of 13.9 pN), the distance between the two traps was fixed and measurements of the hairpin folding were collected by monitoring the separation between the two trapped beads. In order to measure the hairpin dynamics at constant force, we employed a passive force clamp by adjusting the power of one of the optical traps \cite{greenleaf_passive_2005}. Data were sampled at 1 MHz and filtered at the Nyquist frequency.

The point spread function used for deconvolution was measured by preparing a reference construct containing the same DNA handles but without the hairpin between them, and measuring its dynamics under the same conditions as the hairpin. The extension distribution of the reference construct was then fit to a Gaussian function, which was used as the point spread function.

\subsection{\textit{Add} riboswitch aptamer}
We analyzed measurements of the \textit{add} riboswitch aptamer first reported in Ref. \cite{neupane2011single}. Briefly, the aptamer was attached at each end to beads held in optical traps via kb-long duplex handles. Measurements at a constant force of 11.1 pN were done similarly to the hairpin, but with data sampled at 20 kHz. Our analysis used the first 5 s of the trajectory.

\section{Deconvolution}
To deconvolve the point spread function from the hairpin data, we used a modified version of a deconvolution routine previously used on hairpin data \cite{woodside_direct_2006}. The measured extension distribution of the hairpin $p(x)$ and the point spread function $S(x)$ determined from the reference construct were first convolved with a Gaussian function with a standard deviation of 2 $nm$ to smooth the data. We then made the initial guess of the deconvolved distributions $p^{(0)}_{\textrm{deconv}}(x) = p(x)$, and calculated their convolution with the point spread function $p^{(0)}_{\textrm{conv}} = p^{(0)}_{\textrm{deconv}}(x)\otimes S(x)$. We then computed the free energy landscape corresponding to the measured and convolved distributions as $G(x) = -log(p(x))$ and $G^{(0)}_{\textrm{conv}}(x) = -log(p^{(0)}_{\textrm{conv}}(x))$ respectively, subtracting the minimum free energy value from each landscape so that both were strictly positive. We then updated our estimate of the deconvolved distribution iteratively using the difference between the two free energy landscapes:
\begin{align*}
p^{(i+1)}_{\textrm{deconv}}(x) = p^{(i)}_{\textrm{deconv}}(x)\cdot e^{-(G(x)-G^{(i)}_{\textrm{conv}}(x))},
\end{align*}
renormalizing the updated guess of the deconvolved distribution after each iteration. We ran this process for 15 iterations, at which point the convolved and measured distributions matched well but the noise spikes common to deconvolution procedures had not yet appeared.

\setcounter{figure}{0}
\renewcommand{\figurename}{Fig.}
\renewcommand{\thefigure}{S\arabic{figure}}

\begin{figure*}[h!]
    \centering
    \includegraphics[width=0.99\textwidth]{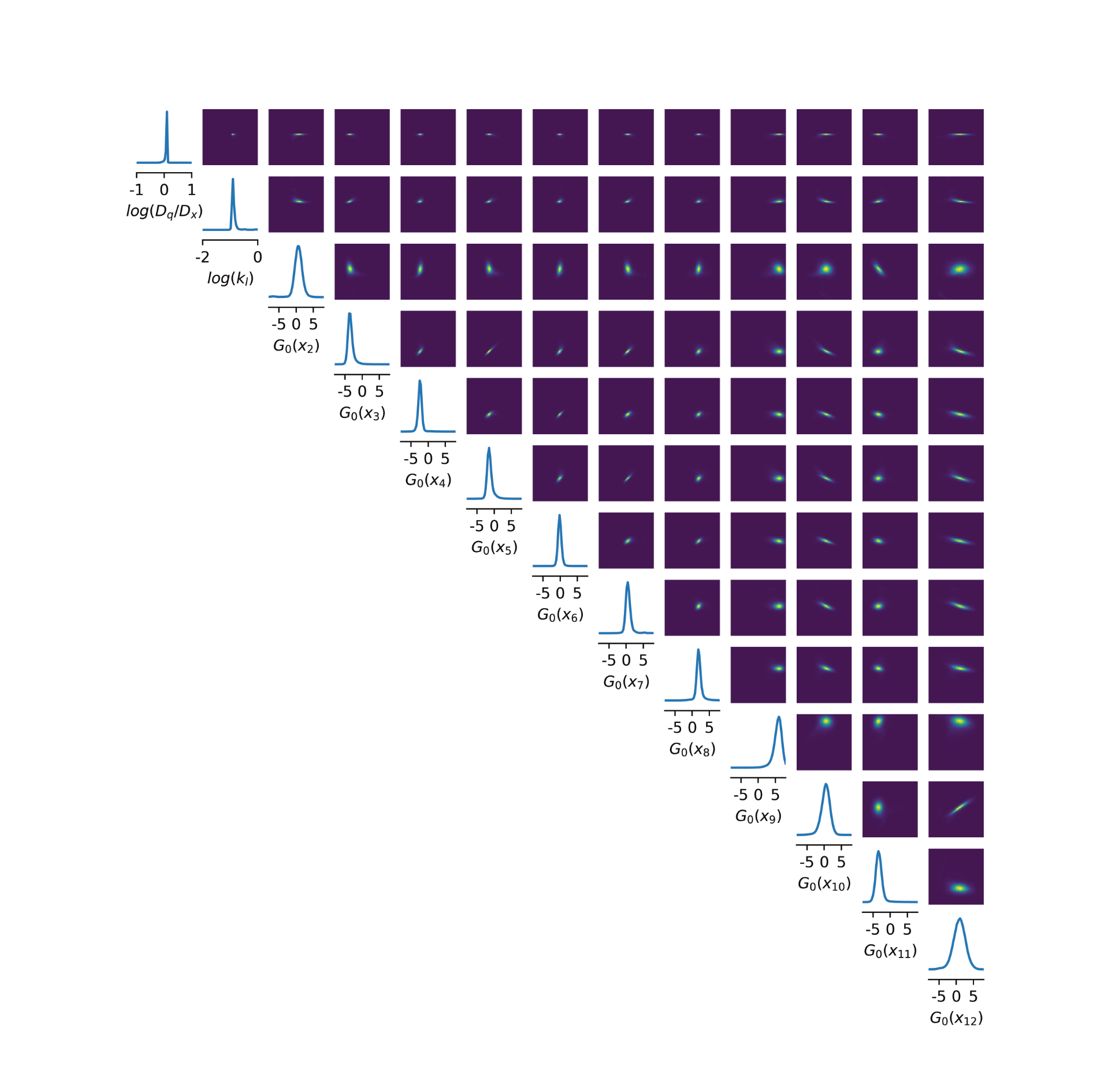}
    \caption{One and two-dimensional marginals of the full posterior distribution $f_{\psi}(\boldsymbol{\theta} | \boldsymbol{q}_{[1:N]}^{\textrm{exp}})$.}
    \label{fig:SI1}
\end{figure*}

\newpage
\begin{figure*}[h!]
    \centering
    \includegraphics[width=0.65\textwidth]{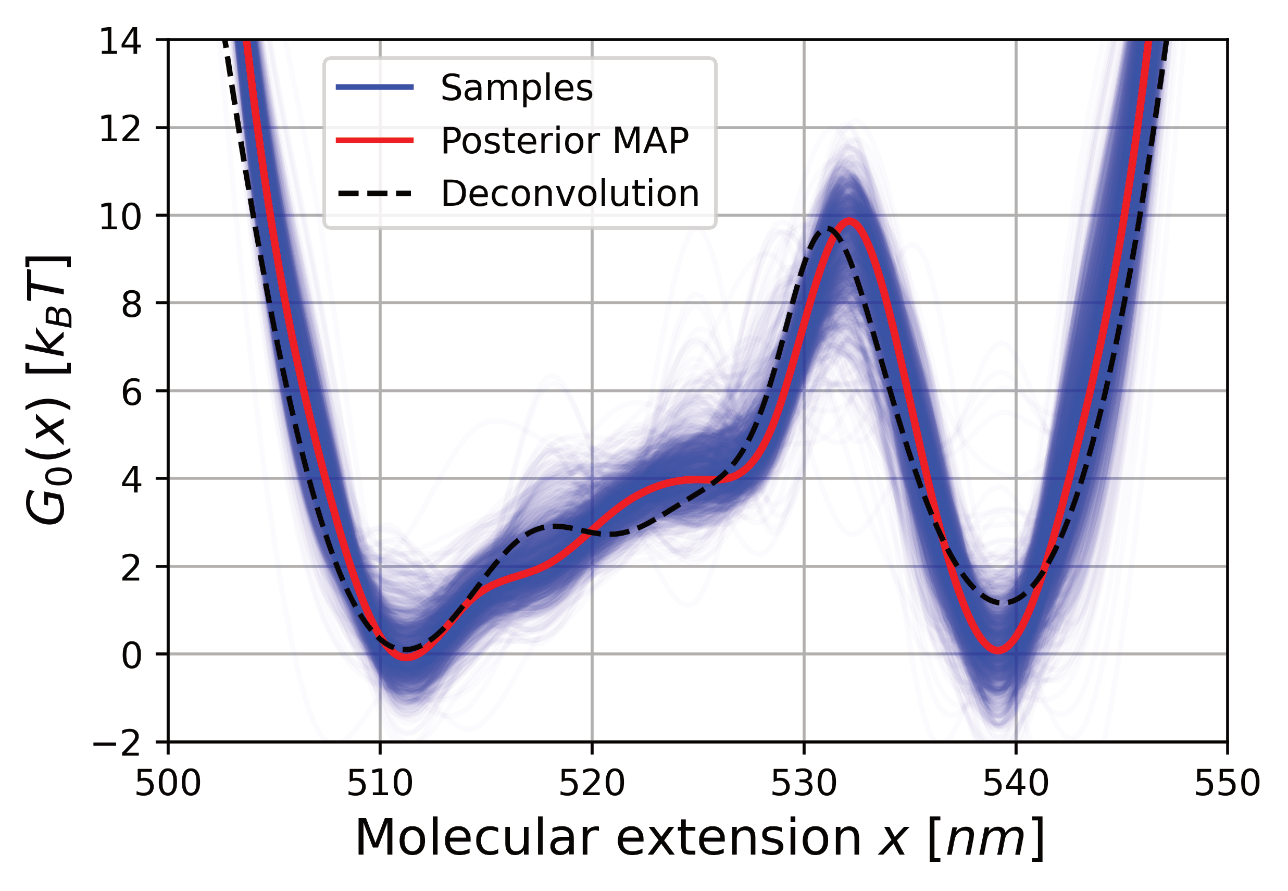}
    \caption{Reconstructed free energy profile. Best estimate $\hat{\boldsymbol{\theta}}^{\textrm{exp}}_{\textrm{MAP}}$ (Maximum a posteriori, MAP) in red, and posterior samples covering a 95 \% confidence interval as blue thin lines. The black line indicates the estimate  using deconvolution.}
    \label{fig:SI2}
\end{figure*}

\newpage
\begin{figure*}[h!]
    \centering
    \includegraphics[width=0.65\textwidth]{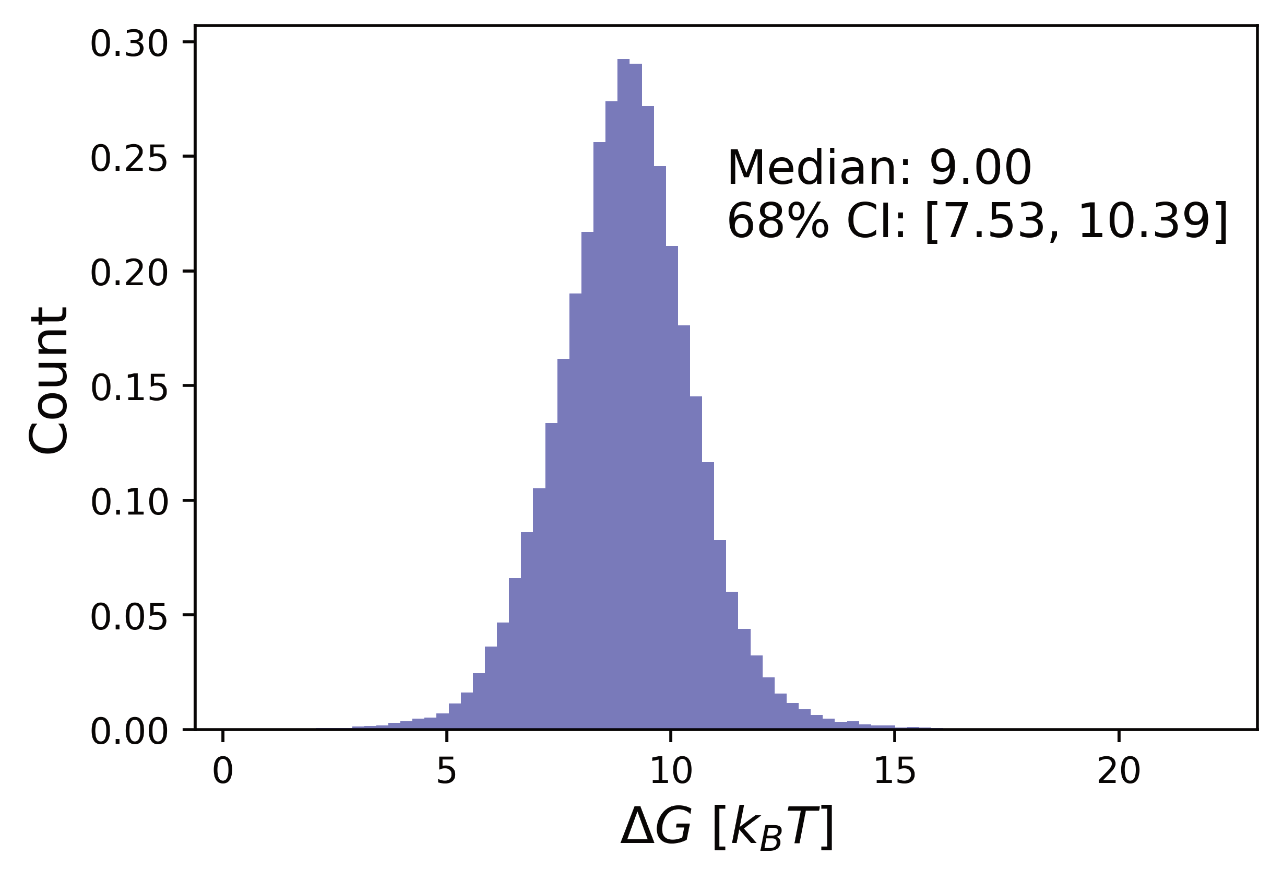}
    \caption{Distribution of barrier heights computed from posterior samples. The Median barrier height is 9.5$k_{\textrm{B}}T$. The lower bound of the 68$\%$ interval is at 8.09 $k_{\textrm{B}}T$ and the upper bound is at 10.72 $k_{\textrm{B}}T$}
    \label{fig:SI3}
\end{figure*}

\newpage
\begin{figure*}[h!]
    \centering
    \includegraphics[width=0.65\textwidth]{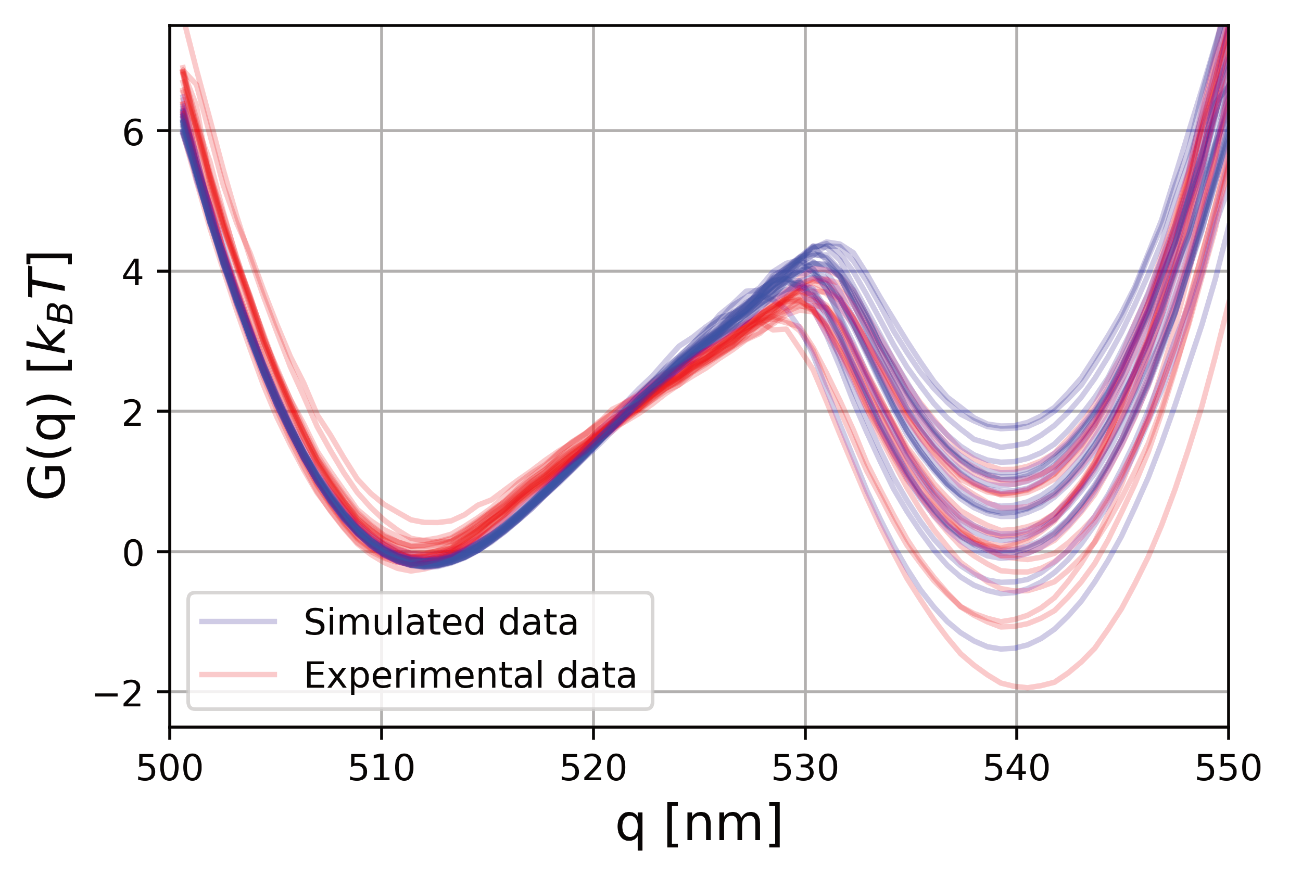}
    \caption{Potential of mean force extracted from 20 individual trajectories (experimental and simulated), which were aligned to minimize the vertical spread between the PMFs.}
    \label{fig:SI4}
\end{figure*}

\newpage
\begin{figure*}[h!]
    \centering
    \includegraphics[width=1\textwidth]{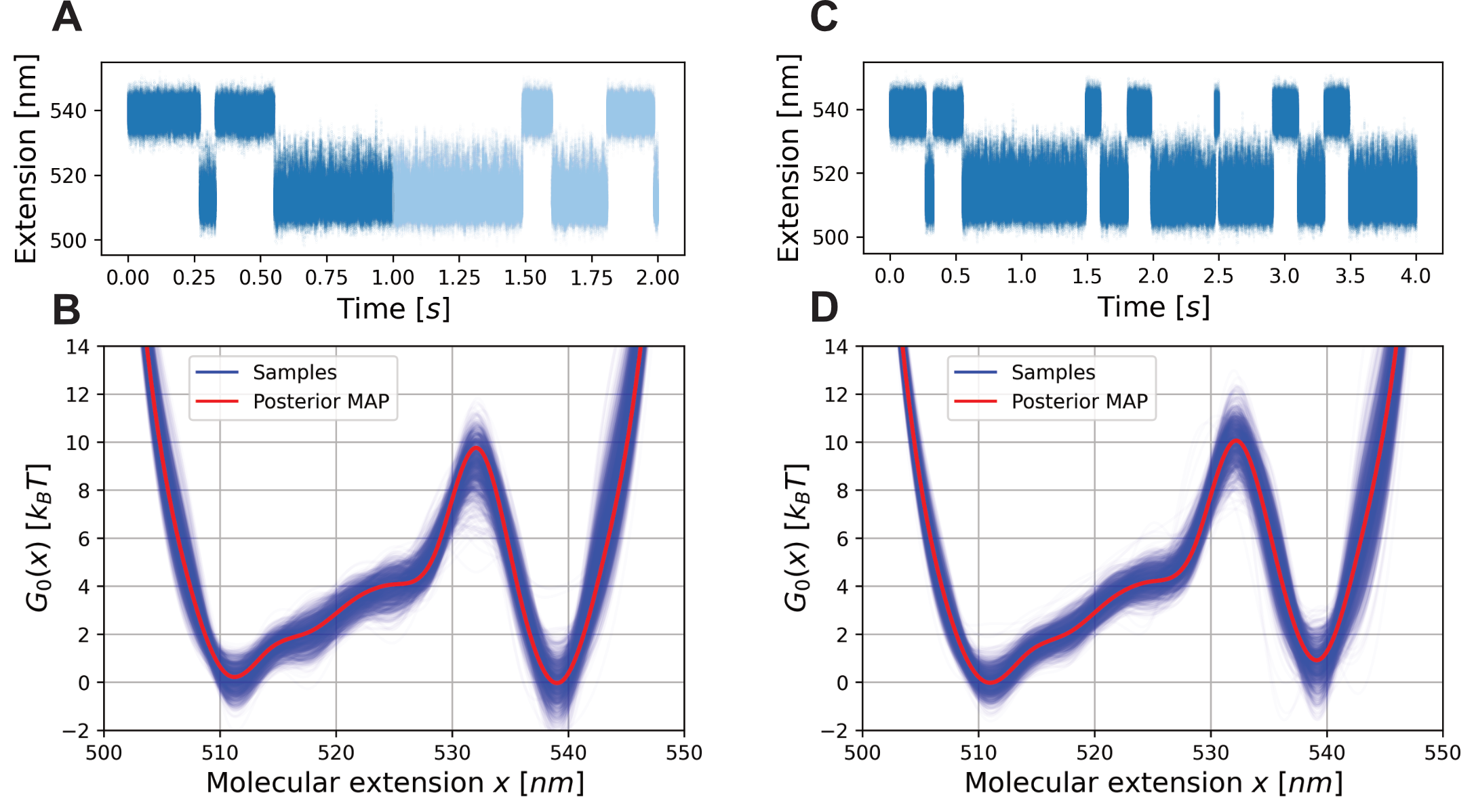}
    \caption{Inference for trajectories of different lengths. (A) One-second-long trajectory. (B) Estimated free-energy landscape inferred from trajectory (A), with the maximum  posteriori (MAP) estimate $\hat{\boldsymbol{\theta}}^{\textrm{exp}}_{\textrm{MAP}}$ shown in red and posterior samples corresponding to the 68\,\% confidence interval shown as thin blue lines. (C) Four-second-long trajectory obtained by concatenating two two-second-long trajectories. (D) Estimated free-energy landscape inferred from trajectory (C), with the MAP estimate shown in red and posterior samples spanning the 68\,\% confidence interval shown as thin blue lines.}
    \label{fig:SI5}
\end{figure*}

\newpage
\begin{figure*}[h!]
    \centering
    \includegraphics[width=0.65\textwidth]{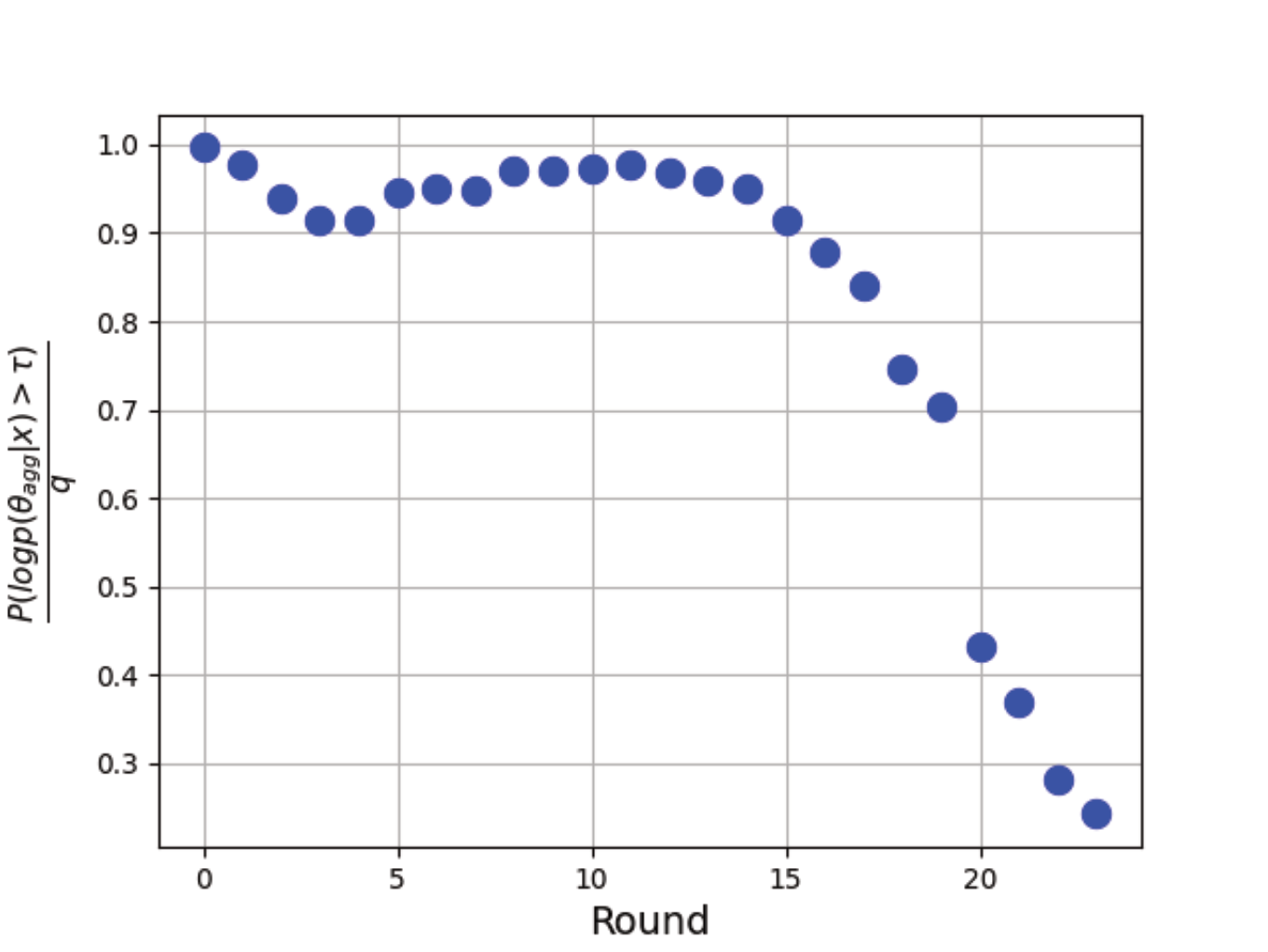}
    \caption{The ratio of the size between the 99$\%$ confidence interval of the posterior for one trajectory versus the size of the 99$\%$ confidence interval from the average posterior over all analyzed trajectories. A value of one means that the posterior 99$\%$ confidence intervals of the aggregated and the individual posterior perfectly overlap. A value smaller than 1 indicates that only a part of the aggregate posterior is covered by the single posterior, indicating that the posteriors are slightly different.
}
    \label{fig:SI6}
\end{figure*}

\newpage
\begin{figure*}[h!]
    \centering
    \includegraphics[width=1\textwidth]{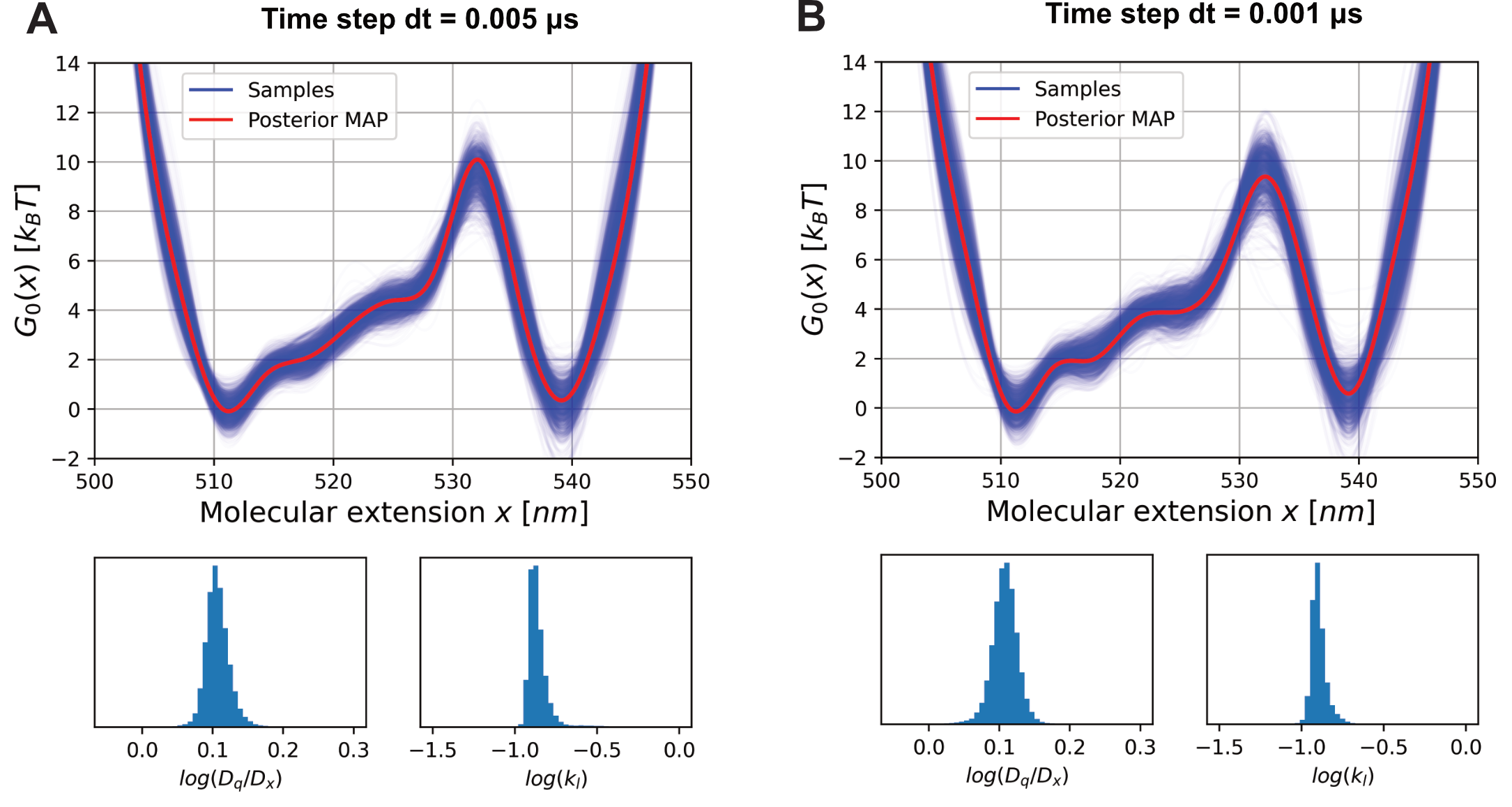}
    \caption{Inference runs with different time discretizations. (A) Inference performed with the simulation time step reduced by a factor of 2. (B) Inference performed with the simulation time step reduced by a factor of 10. All other parameters were identical to those used in the results discussed in the section “Hyperparameters.”}
    \label{fig:SI7}
\end{figure*}

\newpage
\begin{figure*}[h!]
    \centering
    \includegraphics[width=1\textwidth]{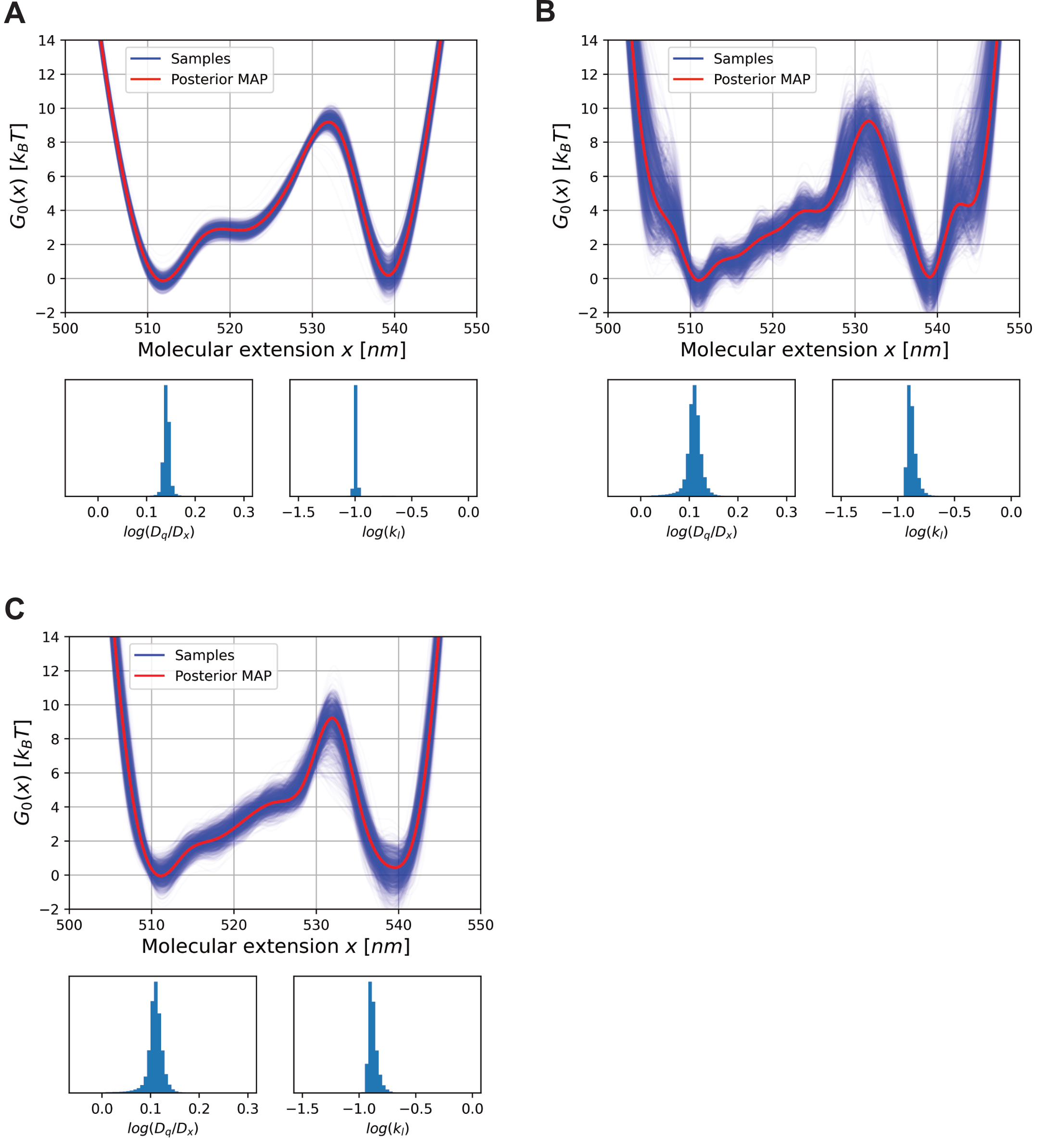}
    \caption{Inference runs with different spline hyperparameters. (A) Inference performed with 10 spline nodes. (B) Inference performed with 20 spline nodes. (C) Inference performed with the second to last and last node offset by 20 and 50 $k_{\textrm{B}}T$ respectively. All other parameters were identical to those used in the results discussed in the section “Hyperparameters.”}
    \label{fig:SI8}
\end{figure*}

\newpage
\begin{figure*}[h!]
    \centering
    \includegraphics[width=0.5\textwidth]{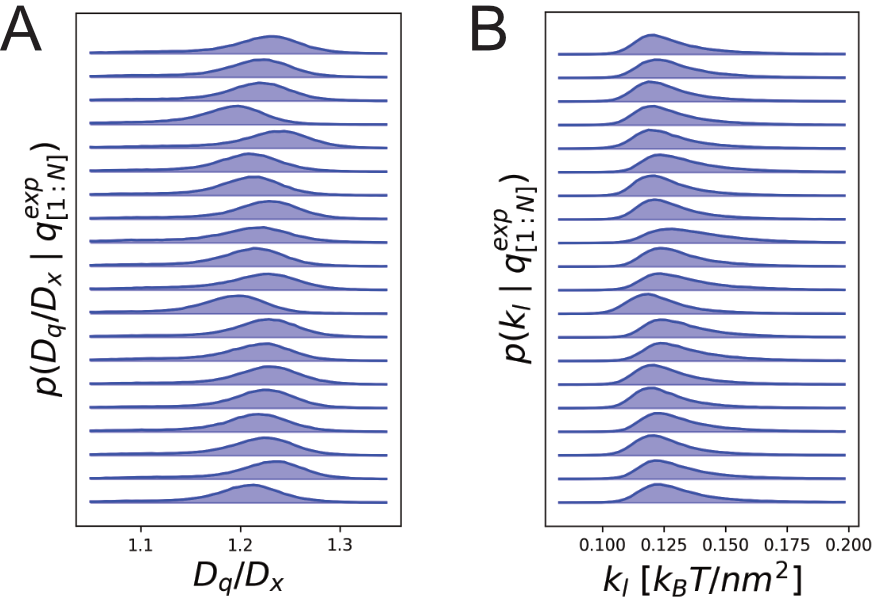}
    \caption{Diffusion coefficients and linker stiffness estimates for synthetic data. Marginal posterior distributions obtained using 20 independent synthetic trajectories, quantifying the inference on (A) the ratio of diffusion coefficients $D_q / D_x$, and (B) the linker stiffness $k_l$.}
    \label{fig:SI9}
\end{figure*}

\newpage
\begin{figure*}[h!]
    \centering
    \includegraphics[width=1\textwidth]{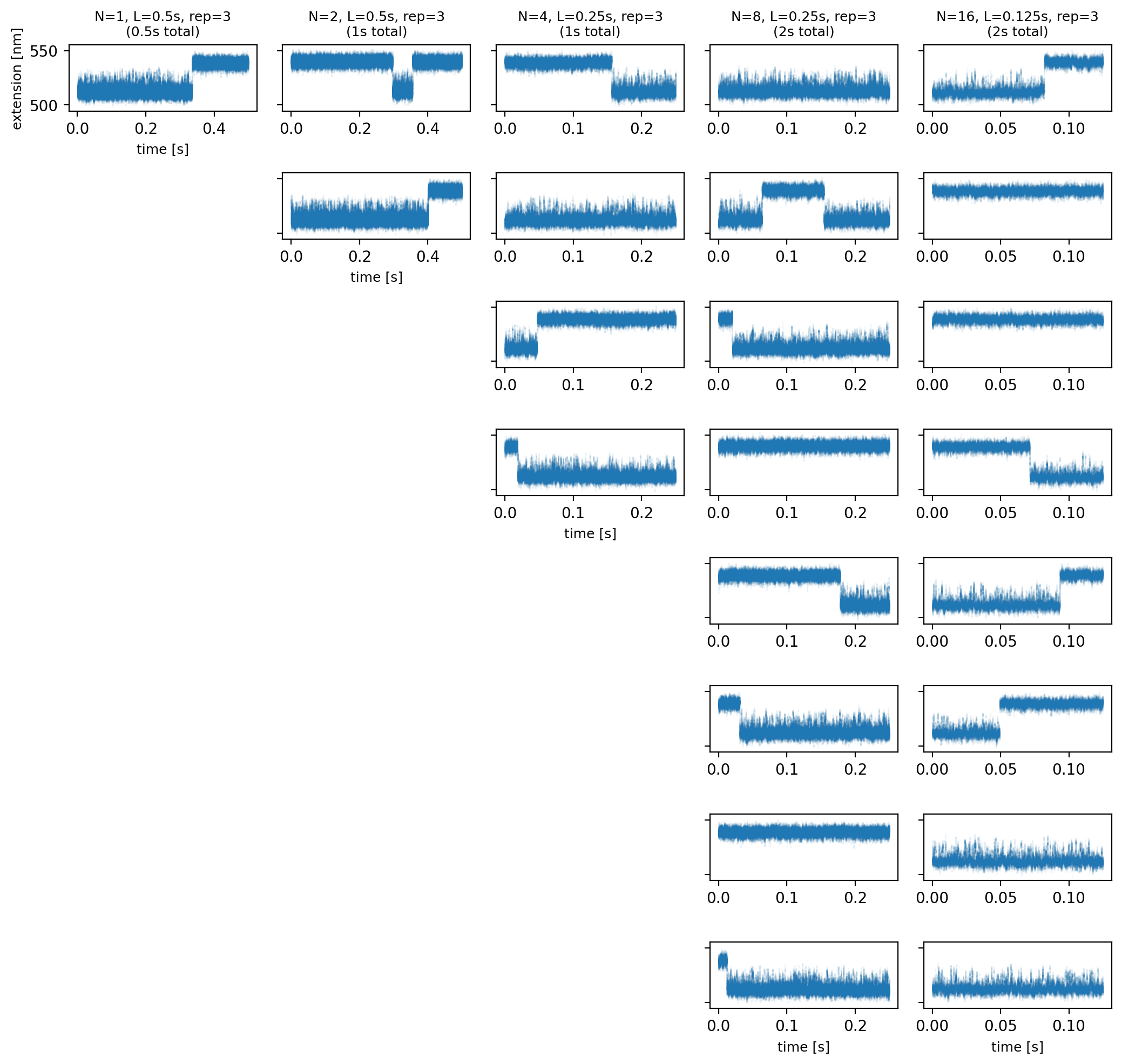}
    \caption{Examples of randomly cropped segments for representative $(N, L)$ settings. Each column corresponds to a different combination of segment number $N$ and segment length $L$. Each row shows the individual segment drawn from one of the 20 experimental trajectories.}
    \label{fig:SI10}
\end{figure*}

\newpage
\begin{figure*}[h!]
    \centering
    \includegraphics[width=1\textwidth]{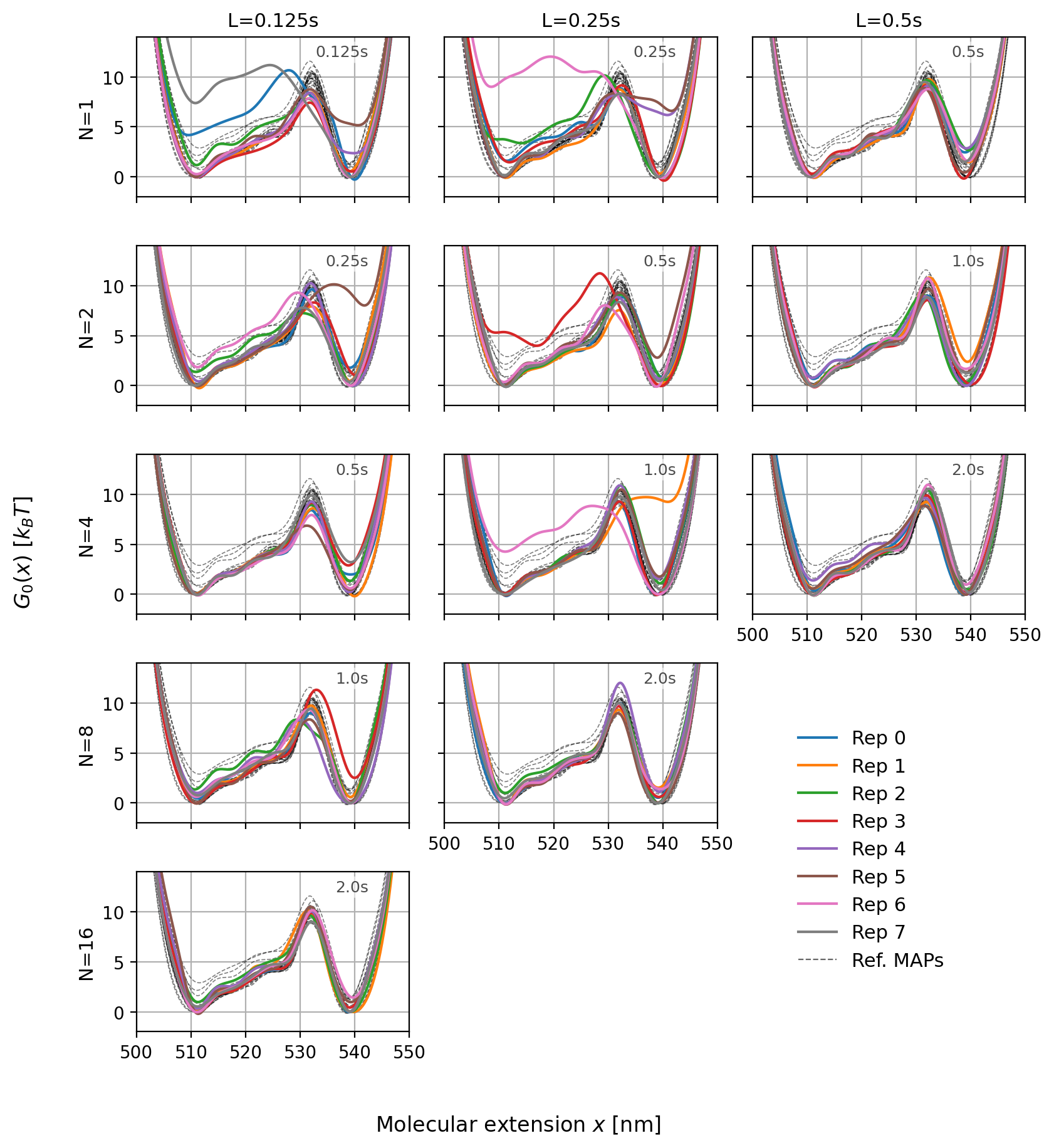}
    \caption{Inference results for combined trajectory segments. Each panel shows the free-energy landscape $G_0(x)$ obtained from the maximum a posteriori estimate (MAP) for a given combination of segment number $N$ (rows) and segment length $L$ (columns). The total duration $N \cdot L$ is indicated in the upper right of each panel. Colored solid lines correspond to eight independent replicates (Rep~0--7), each obtained from a different random selection of trajectory segments. For reference, the 20 individual MAP landscapes inferred from the full 2~s trajectories are shown as dashed gray lines in every panel.}
    \label{fig:SI11}
\end{figure*}

\bibliographystyle{unsrt}
\bibliography{main.bib}

@article{nielsen2020generalization,
  title={On a generalization of the Jensen--Shannon divergence and the Jensen--Shannon centroid},
  author={Nielsen, Frank},
  journal={Entropy},
  volume={22},
  number={2},
  pages={221},
  year={2020},
  publisher={MDPI}
}

@article{englesson2021generalized,
  title={Generalized jensen-shannon divergence loss for learning with noisy labels},
  author={Englesson, Erik and Azizpour, Hossein},
  journal={Advances in Neural Information Processing Systems},
  volume={34},
  pages={30284--30297},
  year={2021}
}

@article{neupane2011single,
  title={Single-molecule force spectroscopy of the \textit{add} adenine riboswitch relates folding to regulatory mechanism},
  author={Neupane, Krishna and Yu, Hao and Foster, Daniel AN and Wang, Feng and Woodside, Michael T},
  journal={Nucleic acids research},
  volume={39},
  number={17},
  pages={7677--7687},
  year={2011},
  publisher={Oxford University Press}
}

@article{roth2009structural,
  title={The structural and functional diversity of metabolite-binding riboswitches},
  author={Roth, Adam and Breaker, Ronald R},
  journal={Annual review of biochemistry},
  volume={78},
  number={1},
  pages={305--334},
  year={2009},
  publisher={Annual Reviews}
}

@article{satija_generalized_2019,
	title = {Generalized {Langevin} {Equation} as a {Model} for {Barrier} {Crossing} {Dynamics} in {Biomolecular} {Folding}},
	volume = {123},
	issn = {1520-6106},
	number = {4},
	urldate = {2025-07-31},
	journal = {The Journal of Physical Chemistry B},
	author = {Satija, Rohit and Makarov, Dmitrii E.},
	year = {2019},
	pages = {802--810},
}

@article{satija_transition_2017,
	title = {Transition path times reveal memory effects and anomalous diffusion in the dynamics of protein folding},
	volume = {147},
	issn = {0021-9606},
	number = {15},
	urldate = {2025-07-31},
	journal = {The Journal of Chemical Physics},
	author = {Satija, Rohit and Das, Atanu and Makarov, Dmitrii E.},
	year = {2017},
	pages = {152707},
}

@article{satija_broad_2020,
	title = {Broad distributions of transition-path times are fingerprints of multidimensionality of the underlying free energy landscapes},
	volume = {117},
	number = {44},
	urldate = {2025-07-31},
	journal = {Proceedings of the National Academy of Sciences},
	author = {Satija, Rohit and Berezhkovskii, Alexander M. and Makarov, Dmitrii E.},
	year = {2020},
	pages = {27116--27123},
}

@article{dudko_intrinsic_2006,
	title = {Intrinsic {Rates} and {Activation} {Free} {Energies} from {Single}-{Molecule} {Pulling} {Experiments}},
	volume = {96},
	issn = {0031-9007, 1079-7114},
	number = {10},
	urldate = {2025-07-30},
	journal = {Physical Review Letters},
	author = {Dudko, Olga and Hummer, Gerhard and Szabo, Attila},
	year = {2006},
	pages = {108101},
}

@article{pierse_distinguishing_2017,
	title = {Distinguishing {Signatures} of {Multipathway} {Conformational} {Transitions}},
	volume = {118},
	issn = {0031-9007, 1079-7114},
	number = {8},
	urldate = {2025-07-30},
	journal = {Physical Review Letters},
	author = {Pierse, Christopher A. and Dudko, Olga K.},
	year = {2017},
	pages = {088101},
}

@article{gebhardt2020full,
author = {J. Christof M. Gebhardt  and Thomas Bornschlögl  and Matthias Rief },
title = {Full distance-resolved folding energy landscape of one single protein molecule},
journal = {Proceedings of the National Academy of Sciences},
volume = {107},
number = {5},
pages = {2013-2018},
year = {2010},
}

@article{engel2014prl,
  title = {Reconstructing Folding Energy Landscape Profiles from Nonequilibrium Pulling Curves with an Inverse Weierstrass Integral Transform},
  author = {Engel, Megan C. and Ritchie, Dustin B. and Foster, Daniel A. N. and Beach, Kevin S. D. and Woodside, Michael T.},
  journal = {Physical Review Letters},
  volume = {113},
  issue = {23},
  pages = {238104},
  numpages = {5},
  year = {2014},
}

@article{manuel2015reconstructing,
  title={Reconstructing folding energy landscapes from splitting probability analysis of single-molecule trajectories},
  author={Manuel, Ajay P and Lambert, John and Woodside, Michael T},
  journal={Proceedings of the National Academy of Sciences},
  volume={112},
  number={23},
  pages={7183--7188},
  year={2015},
}

@article{neupane2012transition,
  title={Transition Path Times for Nucleic Acid Folding Determined from Energy-Landscape Analysis of Single-Molecule Trajectories},
  author={Neupane, Krishna and Ritchie, Dustin B and Yu, Hao and Foster, Daniel AN and Wang, Feng and Woodside, Michael T},
  journal={Physical Review Letters},
  volume={109},
  number={6},
  pages={068102},
  year={2012},
}

@article{neupane2015transition,
  title={Transition-path probability as a test of reaction-coordinate quality reveals {DNA} hairpin folding is a one-dimensional diffusive process},
  author={Neupane, Krishna and Manuel, Ajay P and Lambert, John and Woodside, Michael T},
  journal={The Journal of Physical Chemistry Letters},
  volume={6},
  number={6},
  pages={1005--1010},
  year={2015},
}

@book{buchner2005protein,
  title={Protein folding handbook},
  author={Buchner, Johannes and Kiefhaber, Thomas},
  volume={3},
  year={2005},
  publisher={Wiley-VCH Weinheim}
}

@article{dill2012protein,
  title={The protein-folding problem, 50 years on},
  author={Dill, Ken A and MacCallum, Justin L},
  journal={Science},
  volume={338},
  number={6110},
  pages={1042--1046},
  year={2012},
}

@article{dill_levinthal_1997,
  title={From Levinthal to pathways to funnels},
  author={Dill, Ken A and Chan, Hue Sun},
  journal={Nature Structural Biology},
  volume={4},
  number={1},
  pages={10--19},
  year={1997},
}

@article{petrosyan_single-molecule_2021,
    title = {Single-{Molecule} {Force} {Spectroscopy} of {Protein} {Folding}},
    volume = {433},
    issn = {10898638},
    number = {20},
    journal = {Journal of Molecular Biology},
    author = {Petrosyan, Rafayel and Narayan, Abhishek and Woodside, Michael T.},
    year = {2021},
    pmid = {34418422},
}

@article{woodside_direct_2006,
    title = {Direct measurement of the full, sequence-dependent folding landscape of a nucleic acid},
    volume = {314},
    number = {5801},
    urldate = {2023-01-11},
    journal = {Science},
    author = {Woodside, Michael T. and Anthony, Peter C. and Behnke-Parks, William M. and Larizadeh, Kevan and Herschlag, Daniel and Block, Steven M.},
    year = {2006},
    pages = {1001--1004},
}

@article{cossio_artifacts_2015,
  title={On artifacts in single-molecule force spectroscopy},
  author={Cossio, Pilar and Hummer, Gerhard and Szabo, Attila},
  journal={Proceedings of the National Academy of Sciences},
  volume={112},
  number={46},
  pages={14248--14253},
  year={2015},
}

@article{neuman_single-molecule_2008,
    title = {Single-molecule force spectroscopy: optical tweezers, magnetic tweezers and atomic force microscopy},
    volume = {5},
    issn = {1548-7091},
    number = {6},
    journal = {Nature Methods},
    author = {Neuman, Keir C and Nagy, Attila},
    year = {2008},
    pages = {491--505},
}

@article{neupane_quantifying_2016,
    title = {Quantifying {Instrumental} {Artifacts} in {Folding} {Kinetics} {Measured} by {Single}-{Molecule} {Force} {Spectroscopy}},
    volume = {111},
    issn = {15420086},
    number = {2},
    journal = {Biophysical Journal},
    author = {Neupane, Krishna and Woodside, Michael T.},
    year = {2016},
    pmid = {27369870},
    pages = {283--286},
}

@article{woodside_reconstructing_2014,
  title={Reconstructing folding energy landscapes by single-molecule force spectroscopy},
  author={Woodside, Michael T and Block, Steven M},
  journal={Annual Review of Biophysics},
  volume={43},
  pages={19--39},
  year={2014},
}

@article{dingeldein_simulation-based_2022,
    title = {Simulation-based inference of single-molecule force spectroscopy},
    volume = {4},
    issn = {2632-2153},
    number = {2},
    urldate = {2023-05-07},
    journal = {Machine Learning: Science and Technology},
    author = {Dingeldein, Lars and Cossio, Pilar and Covino, Roberto},
    year = {2022},
    pages = {025009},
}

@article{cranmer_frontier_2020,
    title = {The frontier of simulation-based inference},
    volume = {117},
    issn = {10916490},
    number = {48},
    journal = {Proceedings of the National Academy of Sciences of the United States of America},
    author = {Cranmer, Kyle and Brehmer, Johann and Louppe, Gilles},
    year = {2020},
    pmid = {32471948},
    pages = {30055--30062},
}

@article{dax_real-time_2021,
    title = {Real-{Time} {Gravitational} {Wave} {Science} with {Neural} {Posterior} {Estimation}},
    volume = {127},
    issn = {0031-9007, 1079-7114},
    number = {24},
    urldate = {2024-07-03},
    journal = {Physical Review Letters},
    author = {Dax, Maximilian and Green, Stephen R. and Gair, Jonathan and Macke, Jakob H. and Buonanno, Alessandra and Schölkopf, Bernhard},
    year = {2021},
    pages = {241103},
}

@article{gao_deep_2024,
  title={Deep inverse modeling reveals dynamic-dependent invariances in neural circuit mechanisms},
  author={Gao, Richard and Deistler, Michael and Schulz, Auguste and Gon{\c{c}}alves, Pedro J and Macke, Jakob H},
  journal={Biorxiv},
  year={2024},
  publisher={Cold Spring Harbor Laboratory}
}

@inproceedings{lueckmann_benchmarking_2021,
  title={Benchmarking simulation-based inference},
  author={Lueckmann, Jan-Matthis and Boelts, Jan and Greenberg, David and Goncalves, Pedro and Macke, Jakob},
  booktitle={International Conference on Artificial Intelligence and Statistics},
  pages={343--351},
  year={2021},
}

@article{lueckmann_flexible_nodate,
  title={Flexible statistical inference for mechanistic models of neural dynamics},
  author={Lueckmann, Jan-Matthis and Goncalves, Pedro J and Bassetto, Giacomo and {\"O}cal, Kaan and Nonnenmacher, Marcel and Macke, Jakob H},
  journal={Advances in Neural Information Processing Systems},
  volume={30},
  year={2017}
}

@article{regaldo-saint_blancard_galaxy_2024,
  title={Galaxy clustering analysis with SimBIG and the wavelet scattering transform},
  author={R{\'e}galdo-Saint Blancard, Bruno and Hahn, ChangHoon and Ho, Shirley and Hou, Jiamin and Lemos, Pablo and Massara, Elena and Modi, Chirag and Dizgah, Azadeh Moradinezhad and Parker, Liam and Yao, Yuling and others},
  journal={Physical Review D},
  volume={109},
  number={8},
  pages={083535},
  year={2024},
}

@article{liphardt_reversible_2001,
  title={Reversible unfolding of single {RNA} molecules by mechanical force},
  author={Liphardt, Jan and Onoa, Bibiana and Smith, Steven B and Tinoco Jr, Ignacio and Bustamante, Carlos},
  journal={Science},
  volume={292},
  number={5517},
  pages={733--737},
  year={2001},
}

@article{covino_2019,
    author = {Covino, Roberto and Woodside, Michael T. and Hummer, Gerhard and Szabo, Attila and Cossio, Pilar},
    title = {Molecular free energy profiles from force spectroscopy experiments by inversion of observed committors},
    journal = {The Journal of Chemical Physics},
    volume = {151},
    number = {15},
    pages = {154115},
    year = {2019},
    issn = {0021-9606},
}

@article{dudko_theory_2008,
  title={Theory, analysis, and interpretation of single-molecule force spectroscopy experiments},
  author={Dudko, Olga K and Hummer, Gerhard and Szabo, Attila},
  journal={Proceedings of the National Academy of Sciences},
  volume={105},
  number={41},
  pages={15755--15760},
  year={2008},
}

@article{hummer_free_2010,
    title = {Free energy profiles from single-molecule pulling experiments},
    volume = {107},
    issn = {00278424},
    number = {50},
    urldate = {2022-08-15},
    journal = {Proceedings of the National Academy of Sciences of the United States of America},
    author = {Hummer, Gerhard and Szabo, Attila},
    year = {2010},
    pages = {21441--21446},
}

@article{boelts_sbi_2024,
    doi = {10.21105/joss.07754},
    url = {https://doi.org/10.21105/joss.07754},
    year = {2025},
    volume = {10},
    number = {108},
    pages = {7754},
    author = {Boelts, Jan and Deistler, Michael and Gloeckler, Manuel and Tejero-Cantero, Alvaro and Lueckmann, Jan-Matthis and Moss, Guy and Steinbach, Peter and Moreau, Thomas and Muratore, Fabio and Linhart, Julia and Durkan, Conor and Vetter, Julius and Miller, Benjamin Kurt and Herold, Maternus and Ziaeemehr, Abolfazl and Pals, Matthijs and Gruner, Theo and Bischoff, Sebastian and Krouglova, Nastya and Gao, Richard and Lappalainen, Janne K. and Mucsányi, Bálint and Pei, Felix and Schulz, Auguste and Stefanidi, Zinovia and Rodrigues, Pedro and Schröder, Cornelius and Zaid, Faried Abu and Beck, Jonas and Kapoor, Jaivardhan and Greenberg, David S. and Gonçalves, Pedro J. and Macke, Jakob H},
    title = {sbi reloaded: a toolkit for simulation-based inference workflows},
    journal = {Journal of Open Source Software}
}

@article{durkan_neural_2019,
  title={Neural spline flows},
  author={Durkan, Conor and Bekasov, Artur and Murray, Iain and Papamakarios, George},
  journal={Advances in Neural Information Processing Systems},
  volume={32},
  year={2019}
}

@inproceedings{greenberg_automatic_nodate,
  title={Automatic posterior transformation for likelihood-free inference},
  author={Greenberg, David and Nonnenmacher, Marcel and Macke, Jakob},
  booktitle={International Conference on Machine Learning},
  pages={2404--2414},
  year={2019},
}

@article{varani_exceptionally_1995,
    title = {Exceptionally stable nucleic acid hairpins},
    volume = {24},
    issn = {1056-8700},
    journal = {Annual Review of Biophysics and Biomolecular Structure},
    author = {Varani, G.},
    year = {1995},
    pages = {379--404},
}

@article{davydova_bacteriophage_2007,
    title = {Bacteriophage {N4} virion {RNA} polymerase interaction with its promoter {DNA} hairpin},
    volume = {104},
    issn = {0027-8424},
    number = {17},
    urldate = {2025-02-14},
    journal = {Proceedings of the National Academy of Sciences of the United States of America},
    author = {Davydova, Elena K. and Santangelo, Thomas J. and Rothman-Denes, Lucia B.},
    year = {2007},
    pages = {7033--7038},
}

@article{papamakarios_fast_2018,
  title={Fast $\varepsilon$-free inference of simulation models with bayesian conditional density estimation},
  author={Papamakarios, George and Murray, Iain},
  journal={Advances in Neural Information Processing Systems},
  volume={29},
  year={2016}
}

@inproceedings{papamakarios_sequential_2019,
  title={Sequential neural likelihood: Fast likelihood-free inference with autoregressive flows},
  author={Papamakarios, George and Sterratt, David and Murray, Iain},
  booktitle={The 22nd International Conference on Artificial Intelligence and Statistics},
  pages={837--848},
  year={2019},
}

@inproceedings{durkan_contrastive_2020,
  title={On contrastive learning for likelihood-free inference},
  author={Durkan, Conor and Murray, Iain and Papamakarios, George},
  booktitle={International Conference on Machine Learning},
  pages={2771--2781},
  year={2020},
}

@article{pyo_memory_2019,
    title = {Memory effects in single-molecule force spectroscopy measurements of biomolecular folding},
    volume = {21},
    issn = {1463-9084},
    number = {44},
    urldate = {2023-03-13},
    journal = {Physical Chemistry Chemical Physics},
    author = {Pyo, Andrew G. T. and Woodside, Michael T.},
    year = {2019},
    pages = {24527--24534},
}

@article{greenleaf_passive_2005,
    title = {Passive {All}-{Optical} {Force} {Clamp} for {High}-{Resolution} {Laser} {Trapping}},
    volume = {95},
    number = {20},
    urldate = {2025-02-25},
    journal = {Physical Review Letters},
    author = {Greenleaf, William J. and Woodside, Michael T. and Abbondanzieri, Elio A. and Block, Steven M.},
    year = {2005},
    pages = {208102},
}

@article{lyons_quantifying_2024,
    title = {Quantifying the {Properties} of {Nonproductive} {Attempts} at {Thermally} {Activated} {Energy}-{Barrier} {Crossing} through {Direct} {Observation}},
    volume = {14},
    number = {1},
    urldate = {2025-02-25},
    journal = {Physical Review X},
    author = {Lyons, Aaron and Devi, Anita and Hoffer, Noel Q. and Woodside, Michael T.},
    year = {2024},
    pages = {011017},
}

@article{gupta_experimental_2011,
    title = {Experimental validation of free-energy-landscape reconstruction from non-equilibrium single-molecule force spectroscopy measurements},
    volume = {7},
    issn = {1745-2481},
    number = {8},
    urldate = {2025-02-25},
    journal = {Nature Physics},
    author = {Gupta, Amar Nath and Vincent, Abhilash and Neupane, Krishna and Yu, Hao and Wang, Feng and Woodside, Michael T.},
    year = {2011},
    pages = {631--634},
}

@article{walder_high-precision_2018,
    title = {High-{Precision} {Single}-{Molecule} {Characterization} of the {Folding} of an {HIV} {RNA} {Hairpin} by {Atomic} {Force} {Microscopy}},
    volume = {18},
    issn = {1530-6984},
    number = {10},
    urldate = {2025-02-25},
    journal = {Nano Letters},
    author = {Walder, Robert and Van Patten, William J. and Ritchie, Dustin B. and Montange, Rebecca K. and Miller, Ty W. and Woodside, Michael T. and Perkins, Thomas T.},
    year = {2018},
    pages = {6318--6325},
}

@article{dingeldein_simulation-based_2025,
    title = {Simulation-based inference of single-molecule experiments},
    volume = {91},
    issn = {0959-440X},
    urldate = {2025-02-27},
    journal = {Current Opinion in Structural Biology},
    author = {Dingeldein, Lars and Cossio, Pilar and Covino, Roberto},
    year = {2025},
    pages = {102988},
}

@article{hummer_free_2001,
    title = {Free energy reconstruction from nonequilibrium single-molecule pulling experiments},
    volume = {98},
    issn = {0027-8424},
    number = {7},
    journal = {Proceedings of the National Academy of Sciences},
    author = {Hummer, Gerhard and Szabo, Atilla},
    year = {2001},
    pages = {3658--3661},
}

@article{hinczewski_mechanical_2013,
    title = {From mechanical folding trajectories to intrinsic energy landscapes of biopolymers},
    volume = {110},
    issn = {00278424},
    number = {12},
    journal = {Proceedings of the National Academy of Sciences of the United States of America},
    author = {Hinczewski, Michael and Gebhardt, Christof M. and Rief, Matthias and Thirumalai, D.},
    year = {2013},
    pmid = {23487746},
    pages = {4500--4505},
}

@article{neupane_protein_2016,
    title = {Protein folding trajectories can be described quantitatively by one-dimensional diffusion over measured energy landscapes},
    volume = {12},
    issn = {1745-2481},
    number = {7},
    urldate = {2025-02-13},
    journal = {Nature Physics},
    author = {Neupane, Krishna and Manuel, Ajay P. and Woodside, Michael T.},
    year = {2016},
    pages = {700--703},
}

@article{hoffer2021observing,
  title={Observing the base-by-base search for native structure along transition paths during the folding of single nucleic acid hairpins},
  author={Hoffer, Noel Q and Neupane, Krishna and Woodside, Michael T},
  journal={Proceedings of the National Academy of Sciences},
  volume={118},
  number={49},
  pages={e2101006118},
  year={2021},
  publisher={National Academy of Sciences}
}


\end{document}